# Emerging quantum critical phase in a cluster spin-glass

Fang Zhang[1, 9, #], Tao Feng[1, #], Yurong Ruan[1], Xiaoyuan Ye[3], Bing Wen[1, 4], Liang Zhou[2], Minglin He[5], Zhaotong Zhuang[6], Liusuo Wu[2], Hongtao He[2], Peijie Sun[6], Zhiyang Yu[3, *], Weishu Liu[1, *] & Wenqing Zhang[1, 7, 8, *]

[1]Department of Materials Science and Engineering, Southern University of Science and Technology, Shenzhen 518055, China.

[2]Department of Physics, Southern University of Science and Technology, Shenzhen 518055, China.

[3]State Key Laboratory of Photocatalysis on Energy and Environment, College of Chemistry, Fuzhou University, Fuzhou 350108, China.

[4]Department of Physics, Hongkong University, Hongkong 999077, China.

[5]Department of Mechanical and Energy Engineering, Southern University of Science and Technology, Shenzhen 518055, China.

[6]Beijing National Laboratory for Condensed Matter Physics, Institute of Physics, Chinese Academy of Sciences, Beijing 100190, China.

[7]Shenzhen Institute for Quantum Science and Engineering, Southern University of Science and Technology, Shenzhen 518055, China.

[8]Shenzhen Municipal Key-Lab for Advanced Quantum Materials and Devices & Guangdong Provincial Key Lab for Computational Science and Materials Design, Southern University of Science and Technology, Shenzhen 518055, China.

[9]Present address: Department of Physics, University of California at Berkeley, Berkeley 94720, USA.

[#]These authors contributed equally to this work.

*Corresponding authors: yuzyemlab@fzu.edu.cn, liuws@sustech.edu.cn; zhangwq@sustech.edu.cn

**Abstract:**

**Magnetic frustration has been recognized as pivotal to investigating new phases of matter in correlation-driven Kondo breakdown quantum phase transitions that are not clearly associated with broken symmetry. The nature of these new phases, however, remains underexplored. Here, we report quantum criticalities emerging from a cluster spin-glass in the heavy-fermion metal $TiFe_xCu_{2x-1}Sb$, where frustration originates from intrinsic disorder. Specific heat and magnetic Grüneisen parameter measurements under varying magnetic fields exhibit quantum critical scaling, indicating a quantum critical point near 0.13 Tesla. As the magnetic field increases, the cluster spin-glass phase is progressively suppressed. Upon crossing the quantum critical point, resistivity and Hall effect measurements reveal enhanced screening of local moments and an expanding Fermi surface, consistent with the Kondo breakdown scenario.**

Magnetic frustration in antiferromagnetically insulators can induce quantum spin fluctuations that suppress long-range magnetic order (LRMO), leading to highly spin-correlated ground states found in exotic materials such as quantum spin-liquids and spin-glasses (*1, 2*). Magnetic frustration can also occur in heavy-fermion metals, where it interacts with the Ruderman-Kittel-Kasuya-Yosida (RKKY) magnetic interactions among local moments, as well as their magnetic screening by conduction electrons, a phenomenon known as the Kondo effect (*3*). When the interplay between these competing forces and associated quantum fluctuations becomes prominent, it can lead to novel quantum critical phases and behaviours. By exploring different types of quantum phase transitions (QPTs) and their evolution, we can gain unprecedented insights into the origins of strange-metal quantum criticality, one of the most pressing questions in many-body physics (*4*).

Recent studies on QPTs, particularly those transitioning from a long-range antiferromagnetic (AFM) or ferromagnetic (FM) order to a paramagnetic heavy-fermion liquid, have advanced modern theories of quantum criticality (*5-7*). This emerging framework emphasizes that the behaviour of local moments is as critical as the long-wavelength fluctuations of the order parameter in the canonical Landau paradigm (*8, 9*). A crucial corollary of this theory, known as the Kondo breakdown scenario, is that QPTs can arise from the transition of spin-glasses to a heavy fermion-liquid with an enlarged Fermi surface volume (*10*). Extensive theoretical investigations have pointed out that magnetically frustrated systems, featuring both RKKY and Kondo interactions, may undergo such QPTs through varying non-thermal control parameters, such as magnetic field, pressure, or atomic substitution (*11-16*).

$LiV_2O_4$ is an early candidate proposed to host these intriguing QPTs, exhibiting magnetic frustration due to the varied valences of two vanadium atoms. Despite the potential logarithmic temperature-dependence of its specific heat (*17*)—a signature of the onset of the quantum critical phase (*18*), and the observation of a spin-glass state upon Zn-doping (*19*), subsequent studies have shown that the Kondo effect in $LiV_2O_4$ is negligible. Instead, its heavy-fermion behaviours primarily arises from the lightly doped Mott insulator nature of the strongly correlated $a_{1g}$ band (*20*). While other materials, highlighted by $Pr_2Ir_2O_7$ (*21*), Ge-substituted $YbRh_2Si_2$ (*22*), and CePdAl (*23*), have been investigated as potential candidates, conclusive experimental evidence linking spin-correlated quantum states to QPTs remains elusive, particularly in relation to a spin-glass state.

In this study, we report a new family of iron-based heavy-fermion metals, $TiFe_xCu_{2x-1}Sb$, exemplified by $x = 0.70$, where magnetic frustration originates from inherent disorder within the Fe and Cu sites. Through detailed ultra-low temperature thermodynamic measurements, including specific heat and magnetic Grüneisen parameter, we observe a magnetic field-induced quantum critical point (QCP). Upon crossing this QCP, Hall measurements show an increase in the Fermi surface volume consistent with the Kondo breakdown scenario, as partially supported by resistivity measurements that suggest enhanced screening of local moments. Additionally, direct current (dc) and alternating current (ac) magnetic susceptibility measurements identify a cluster spin-glass ground state where magnetic entities are spin clusters, not isolated atomic spins, which are progressively suppressed with increasing magnetic field.

## Crystal structure and disorder-induced local moments

Figure 1A shows the schematic crystal structure of TiFe$_{0.7}$Cu$_{0.4}$Sb, an off-stoichiometric compound with Heusler-like characteristics (*24*) that belongs to the $F\bar{4}3m$ (No. 216) space group. In this structure, Ti and Sb atoms form the skeleton of a face-centered cubic lattice, while Fe and Cu atoms are expected to partially occupy both the 4$c$ and 4$d$ Wyckoff sites in a disordered manner (*25*). The uniform distribution and expected atomic arrangements of the atoms are confirmed through the integrated differential phase contrast (iDPC) imaging, depicted in Fig. 1B (also see fig. S2). Specifically, energy-dispersive spectroscopy (EDS) maps for Fe (Fig. 1C) and Cu (Fig. 1D) reveal their random spatial distribution, indicating significant chemical disorder between these two competing elements across the 4c and 4d sites. In contrast, neighboring Ti and Sb atoms exhibit periodic arrangements, as shown in the inset of Fig. 1B.

This unique crystal structure of TiFe$_{0.7}$Cu$_{0.4}$Sb leads to randomly distributed local moments, primarily originating from the partially filled $e_g$ orbitals of Fe atoms (fig. S5). Due to disorder-induced variations in the crystal field environment around Fe atoms at different sites, their magnetic moments show considerable variation, ranging from 0.0 to 1.5 $\mu_B$, as demonstrated by our *ab initio* calculations (see Methods). This behaviour can be attributed to Fe being generally non-magnetic in a half-Heusler structural environment or weakly magnetic in a full-Heusler environment, both of which can coexist in such a disordered system. Low-temperature magnetization measurements, $M(\mu_0 H)$, reveal a saturation moment of approximately 0.05 $\mu_B$ per Fe atom (fig. S4), suggesting that most Fe atoms possess negligible magnetic moments. This combination of structural disorder and small average magnetic moments identifies TiFe$_{0.7}$Cu$_{0.4}$Sb as a dilute Kondo disordered system (*26*), characterized by magnetic ions randomly dispersed throughout the lattice.

## Thermodynamic evidence for a magnetic field-induced QCP

We conducted specific heat measurements on single-phase polycrystalline TiFe$_{0.7}$Cu$_{0.4}$Sb samples (fig. S1), with results shown in Fig. 2A-C. To isolate the electronic specific heat coefficient, $C_{el}(T)$, we analyzed the total specific heat, $C_{tot}(T)$, which consists of contributions from nuclear ($C_{nuc}$), magnetic Schottky ($C_{sch}$), phonon ($C_{ph}$), and electronic ($C_{el}$) components (see Methods for further details). Figure 2A shows that $C_{el}(T)/T$ transitions from a $\log(1/T)$-dependent behaviour at high temperatures to either a constant value (at high fields) or an approach toward a constant value (at low fields) upon cooling. This transition is characterized by a crossover temperature scale, $T_{cs}$, which decreases with the applied fields for $\mu_0 H < \mu_0 H_c$, but increases for $\mu_0 H > \mu_0 H_c$, where the critical field is approximately $\mu_0 H_c \approx 0.13$ T (Fig. 3C). As a result, a fan-shaped critical regime emerges, in which the electronic specific heat $C_{el}(T)/T$ follows a logarithmic dependence on temperature. The power-law divergence of $C_{el}(T)$ at $\mu_0 H_c \approx 0.13$ T exhibits a logarithmic dependence, $C_{el}(T)/T \propto \log(T_0/T)$, over a broad temperature range (0.08 K to 2 K), with $T_0 = 7.68(1)$ K being the characteristic temperature associated with spin fluctuation energies. This feature, as observed over an extended temperature range, is widely regarded as a hallmark of quantum criticality (*27*).

Further support for a magnetic field-induced QCP is provided by the Sommerfeld coefficient $\gamma(T)$ as $T \to 0$, where $\gamma(T) = C_{el}(T)/T$. This coefficient is expected to diverge at the critical field $\mu_0 H_c$. As shown in Fig. 2B, $\gamma$ (calculated at our lowest measured temperature of $T = 64$ mK) initially increases slowly with applied magnetic field, reaching a broad maximum of 67.3 mJ mole-Fe$^{-1}$ K$^{-2}$ near the critical field $\mu_0 H_c \approx 0.13$ T, before sharply decreasing at higher fields. Although a clear divergence in $\gamma$ was not observed due to the broad peak, the distinct $\lambda$-shaped curve in the $\gamma$-$\mu_0 H$ diagram conforms to the existence of a continuous second-order quantum phase transition at $\mu_0 H_c$ (*18*).

During these measurements, we noticed that the value of $C_{el}(T)/T$ is nearly two orders of magnitude smaller than those typically found in heavy-fermion systems with LRMO, which generally reach several thousand mJ mole$^{-1}$K$^{-2}$ around 0.1 K (*5, 6, 9*). Since the low-temperature quantum critical behaviour is driven by the involvement of local moments, this relatively small $C_{el}(T)/T$ aligns with the unique nature of local moments in this disordered system, where Fe atoms possess varied and predominantly small (even zero at some sites) magnetic moments. Nevertheless, TiFe$_{0.7}$Cu$_{0.4}$Sb still qualifies as a heavy-fermion material, as it demonstrates an effective electronic mass approximately 150 times that of free electrons (see Methods, Sec. 6) and follows to the Kadowaki-Woods relation (see Supplementary Text, Sec. 16)—both characteristic features of heavy-fermion systems.

To further investigate the presence of a QCP, we conducted magnetocaloric effect measurements on the samples. As magnetic fields $\mu_0 H$ was swept down, the sample temperatures were recorded simultaneously (we limited $\mu_0 H$ to a minimum of 0.08 T due to magnetic flux jump noise from the magnets at lower fields). Although the raw data, as shown in Fig. 2D, exhibited large noise, primarily due to the small averaged magnetic moments of the samples (around 0.05 $\mu_B$ per Fe atom), a distinct minimum around $\mu_0 H = 0.10$ T was observed, indicative of strong spin fluctuations and consistent with the presence of a nearby QCP. The magnetic Grüneisen parameter, $\Gamma_B = 1/T(dT/d\mu_0 H)$, was then calculated. This parameter is expected to diverge as any QCP is approached (*28*), following the relationship $\Gamma_B \propto 1/(\mu_0 H - \mu_0 H_c)^{vz}$, where $v$ is the correlation length exponent and $z$ is the dynamical exponent. As shown in Fig. 2E, $\Gamma_B$ initially increases as the magnetic field decreases, reaching a broad maximum near $\mu_0 H \approx 0.17$ T, before decreasing and changing sign at lower fields. By fitting $\Gamma_B$ for $\mu_0 H > \mu_0 H_c$ with $\mu_0 H_c = 0.13$ T, we obtained a value of $vz = 0.33(2)$ [see Supplementary Text, Sec. 11 for more discussion]. This value deviates from the expected $vz = 1$ or 3/2 associated with AFM or FM QPTs under the Hertz-Millis scenario involving symmetry-breaking (*29*), raising questions about the nature of the QCP in disordered TiFe$_{0.7}$Cu$_{0.4}$Sb.

**Transport evidence for Kondo breakdown quantum criticality**

In a dilute Kondo disordered system, Kondo screening occurs independently at each local site, and physical properties are governed primarily by the single-ion characteristic Kondo temperature, $T_K$, which is expected to be extremely low due to disorder ($T_K$ is around 0.5 K in TiFe$_{0.7}$Cu$_{0.4}$Sb; see Supplementary Text, Sec. 12) (*30*). As the temperature decreases below $T_K$, Kondo screening becomes more effective and extends over larger distances, eventually leading to the development of coherent interactions among local magnetic moments. However, understanding how Kondo lattice coherence develops in such disordered systems remains a challenging open question.

In the dilute regime, resistivity $\rho(T)$ exhibits a logarithmic increase with cooling, $\rho(T) \approx -\ln T$, characteristic of the single-ion Kondo effect. Meanwhile, the onset of coherent Kondo scattering causes metallic behaviour, with $\rho(T)$ decreases as the temperature lowers. Consequently, $\rho(T)$ shows a maximum at low temperatures, marking the crossover from incoherent to coherent Kondo scattering, intuitively defined by a coherence temperature, $T_{coh}$. Figure 2F shows $\rho(T)$ for TiFe$_{0.7}$Cu$_{0.4}$Sb under varying magnetic fields (with $T$ limited to a minimum of 0.1 K due to current-induced heating effects; see fig. S3). Similar to the electronic specific heat $C_{el}(T)$, the resistivity $\rho(T)$ is nearly two orders of magnitude smaller than that of other quantum critical materials. At low fields, $\rho(T)$ initially increasing and then decreases as temperature rises, resulting in a maximum in $\rho(T)$, which indicates the onset of Kondo lattice coherence (*31*).

Since the emergence of quantum criticality relates closely to the Kondo coherence, we firstly focused on resistivity data below the maximum value. This restricts our analysis to a narrow temperature range, making detailed scaling analysis difficult. Here, we fitted the data to $\rho(T) = \rho_0 + AT^n$. At zero field, the power coefficient $n$ is approximately 3.6, deviating from the Fermi-liquid behaviour ($n = 2$) seen in the ground state of AFM or FM materials with QCPs. As the field increases, $n$ initially decreases to around 1 at $\mu_0 H = 0.2$ T, and likely increases at higher fields (although $n$ could not be determined for $\mu_0 H > 0.22$ T, as discussed later). The minimum value of $n = 1$ in the $n$-$\mu_0 H$ diagram may indicate a QCP at $\mu_0 H = 0.2$ T or a nearby QCP, considering the significant disorder present in the system (*32*).

With increasing magnetic field, the coherence temperature $T_{coh}$ corresponding to the maximum in $\rho(T)$ shifts monotonically to lower temperature (indicated by the black dashed line in Fig. 2F; see Supplementary Text, Sec. 13 for more quantitative analysis). At $\mu_0 H > 0.22$ T, the maximum peak is unobservable as it shifts below 0.1 K. Similar behaviour is widely observed in other disordered Kondo system [e.g., Ce$_x$La$_{1-x}$Cu$_6$ upon dilution of the magnetic sublattice (*33*)], indicating the destruction of coherent Kondo interactions as tuning the control parameter (magnetic field here). In the Kondo breakdown scenario, applying a magnetic field enhances Kondo screening, allowing more conduction electrons to incorporate with the local moments and form composite fermions. Consequently, local moments are increasingly screened and $T_{coh}$ goes to lower temperatures, consistent with the behaviour observed in TiFe$_{0.7}$Cu$_{0.4}$Sb.

To confirm whether the transport properties align with the Kondo breakdown scenario, we performed Hall measurements, which provide information about the Fermi surface volume. As magnetic field increases, the formation of composite fermions should result in a larger Fermi surface, as more local electrons, originally contributing to local moments, become part of it. Determining the $\rho_H$ in TiFe$_{0.7}$Cu$_{0.4}$Sb is challenging due to strong mixing signals from magnetoresistivity. To address this, we applied the standard anti-symmetrization process to extract $\rho_H$ by sweeping the magnetic fields in both directions (see Methods, Sec. 9; note that data below $\mu_0 H = 0.1$ T are truncated due to magnetic flux jump noise from the magnets). Figure 2G displays several representative isotherms of Hall resistivity $\rho_H$ at low temperatures. Despite the large scattering, $\rho_H$ show a clear evolution from steep to shallow slopes as the magnetic field increases (see black dashed lines in Fig. 2G), indicating changes in the Hall coefficient and, thus, the Fermi surface.

To quantitatively analyze Hall data, we followed the method in Ref. (*34*) and used $R_H^*(\mu_0 H) = R_H^\infty - (R_H^\infty - R_H^0)\gamma(\mu_0 H)$ as a fitting function. Here, $R_H^0$ is the zero-field Hall coefficient and $R_H^\infty$ is the asymptotic differential Hall coefficient at large fields. $\gamma(\mu_0 H)$ is a crossover function parameterized as $\gamma(\mu_0 H) = 1/[1 + (\mu_0 H/\mu_0 H_{cs})^p]$, where $\mu_0 H_{cs}$ is the crossover field and $p$ is a real number. Fits of $\rho_H = \int R_H^*(H) dH$ at different temperatures, and their derivatives corresponding to $R_H^*(\mu_0 H)$, are shown in the solid lines in Fig. 2G and H, respectively. At all measured temperatures, $R_H^*$ decreases monotonically from low-field to high-field. In the single-carrier model of the Hall coefficient with field-independent mobility, $R_H^*$ is inversely proportional to the Fermi surface volume (i.e., carrier density), and thus the observed decrease of $R_H^*$ with increasing magnetic fields supports the Kondo breakdown phenomenon in TiFe$_{0.7}$Cu$_{0.4}$Sb.

**Magnetic evidence for cluster spin-glass ground state**

In most heavy-fermion metals, the onset of a magnetic field-induced QCP of the Kondo breakdown type is typically accompanied by the destruction of LRMO, i.e., the ground state at $T \to 0$ and $\mu_0 H \to 0$ is usually either AFM or FM. However, in TiFe$_{0.7}$Cu$_{0.4}$Sb, due to significant disorder, LRMO is expected to be absent, making the ground state particularly intriguing.

Extensive studies have shown that the random-exchange Heisenberg-Kondo Hamiltonian effectively describes a disordered Kondo lattice (see Supplementary Text, Sec. 10). By extending the spin-rotation symmetry to $SU(M)$ and applying a large-$M$ dynamical mean-field theory, a QPT between a spin-liquid and a heavy-fermion liquid has been identified (*13, 15*), leading to the emergence of a Kondo breakdown QCP. However, when the spin rotation symmetry is broken and magnetic order is associated, this QPT is most likely to occur between a spin-glass and a heavy-fermion liquid, as analyzed by renormalization group (*12*) and quantum Monte Carlo techniques (*11, 16*) within the context of $SU(2)$ symmetry—a more realistic scenario potentially relevant to our findings here.

To investigate whether TiFe$_{0.7}$Cu$_{0.4}$Sb exhibits spin-glass behaviour, we performed low-temperature zero-field-cooled (ZFC) and field-cooled (FC) dc magnetic susceptibility $\chi(T)$ measurements using a custom-built setup (see Methods). As shown in Fig. 3A, the ZFC and FC curves overlap initially when cooling begins at 1.8 K. However, as the temperature decreases, the separation between the ZFC and FC branches become evident, providing compelling evidence for a spin-glass state. The absence of peaks in the curves indicates that the spin-freezing temperature $T_f$ is below 0.4 K.

Additionally, high-temperature ZFC and FC measurements (above 2K, using standard techniques; see

Methods) revealed a peak at $T'_f = 120$ K, suggesting that a spin-glass phase also exists at high temperatures (inset of Fig. 3A). This high-temperature spin-freezing behaviour is typical in disordered systems (1), but is particularly notable in TiFe$_{0.7}$Cu$_{0.4}$Sb as it undergoes a two-stage freezing process of its magnetic moments: the first at $T'_f = 120$ K, followed by a second stage at much lower temperatures, below 0.4 K. To roughly estimate the number of active magnetic moments contributing to the low-temperature quantum critical behaviour, we calculated the low-temperature magnetic entropy, $S_m$, from the magnetic specific heat, $C_m$ (see Methods, Sec. 6) through $S_m = \int_0^T \frac{C_m}{T} dT$, as shown in Fig. 3B. The theoretical magnetic entropy for a spin-$S$ particle is $S_m = R\ln(2S + 1)$, assuming an average spin value of approximately $S \approx 0.05$, derived from saturated magnetization. However, the calculated $S_m$ is only 34.6 mJ mol$^{-1}$ K$^{-1}$, much lower than the theoretical value of 792.5 mJ mol$^{-1}$ K$^{-1}$. This indicates that nearly all magnetic moments freeze around $T'_f = 120$ K, with only about 4.3% actively contributing to the low-temperature quantum critical process.

To further clarify the nature of the low-temperature spin-glass phase, we performed ac susceptibility measurements, allowing us to probe much lower temperatures (down to 0.05 K). As shown in Fig. 3C, the real part of the ac susceptibility $\chi'(T)$ responses differently at various driving frequencies (the change is small due to the small magnetic moments), indicating a broad distribution of the spin relaxation time. Notably, a peak in $\chi'(T)$ is centered at the freezing temperature $T_f = 0.35(5)$ K at 79 Hz, providing a clear signature of the spin-glass phase with short-range correlations below $T_f$. As seen in Fig. 3C, $T_f$ increases while the peak height decreases at higher driving frequencies, which are typical characteristics of spin-glasses. A common criterion for assessing the frequency dependence of $T_f$ is the relative shift in freezing temperature per decade of frequency, defined as $\delta T_f = \Delta T_f/[T_f \Delta \log(f)]$, where $\Delta T_f$ is the change in freezing temperature, and $\Delta \log(f)$ is the change in the logarithm of the frequency. For TiFe$_{0.7}$Cu$_{0.4}$Sb, we found $\delta T_f = 0.04$ for frequencies ranging from 79 to 9984 Hz, an intermediate value between those reported for canonical spin-glass systems ($\delta T_f \approx 0.001$) (35) and for non-interacting superparamagnetic systems ($\delta T_f \approx 0.1$) (36). This $\delta T_f$ value characterizes the ground state as the so-called cluster spin-glass.

In a cluster spin-glass, the magnetic entities are not individual atomic spins but clusters of spins that behave collectively. In TiFe$_{0.7}$Cu$_{0.4}$Sb, we propose the following scenario: the randomly distributed and varied-sized atomic spins group into clusters, which is a common feature in disordered magnetic systems. As the temperature drops to $T'_f = 120$ K, the atomic spins inside each cluster freeze, marking the first stage of spin freezing. Each cluster, however, retains a small but non-zero net magnetic moment. At lower temperatures, these clusters themselves behave as individual magnetic entities and contribute to the quantum critical behaviour. Further cooling below $T_f = 0.35$ K leads to the freezing of these spin clusters, resulting in the formation of a cluster spin-glass ground state as $T \rightarrow 0$.

The characteristic relaxation time of a single spin flip $\tau_0$ can be estimated from the frequency dependence of freezing temperature $T_f$ using $\frac{1}{f} = \tau_0 \left(\frac{T_f - T_{SG}}{T_{SG}}\right)^{-z'v'}$, where $T_{SG}$ is the spin-glass temperature as frequency approaches zero, and $z'v'$ is the dynamic critical exponent (*1, 37*). For individual spin flip processes, $\tau_0$ is typically around $10^{-12}$ to $10^{-13}$ s (*36*). We estimated $\tau_0 = 5.43 \times 10^{-10}$ s by linear fitting $\ln\left(\frac{1}{f}\right) = \ln(\tau_0) - z'v' \ln\left(\frac{T_f - T_{sg}}{T_{sg}}\right)$ as shown in Fig. 3D. The significantly larger $\tau_0$ reflects the spin-flip process in TiFe$_{0.7}$Cu$_{0.4}$Sb is much slower, suggesting the presence of interacting clusters rather than individual spins [note the fitted $z'v' = 7.06$ also aligns with other cluster spin-glass materials (*38, 39*)]. The presence of interacting clusters is further supported by the departure of frequency dependence of $T_f$ from the Arrhenius law (where $T_0 = 0$ for canonical spin-glass) and its fit to the empirical Vogel-Fulcher law, $f = f_0 \exp\left(-\frac{E_a}{k_B(T_f - T_0)}\right)$, where $f_0$ is the characteristic frequency ($f_0 = 1/\tau_0$), $E_a$ is the activation energy, and $T_0$ is the Vogel-Fulcher temperature (*36*). For TiFe$_{0.7}$Cu$_{0.4}$Sb, we obtained $T_0 = 0.28$ K by fitting $T_f = \frac{E_a}{k_B} \frac{1}{\ln(f_0/f)} + T_0$ (see Fig. 3D). A nonzero value of $T_0$ arises from spin interactions and indicates the formation of spin clusters. The proximity of $T_0$ to $T_f$ suggests that the RKKY interaction between clusters is strong, enabling the formation of Kondo coherence at low temperatures (we also obtained a reasonable estimate of the activation energy $E_a \approx 3.49 k_B T_f$).

Having confirmed the cluster spin-glass nature of the ground state, we investigated the influence of applied magnetic fields on the ac susceptibility, $\chi'(T)$. Magnetic fields are expected to suppress the cluster spin-glass phase, leading to a transition to a heavy-fermion liquid phase as the system approaches and crosses the QCP at $\mu_0 H > \mu_0 H_c$. As shown in Fig. 3F, the peak in $\chi'(T)$ gradually diminishes as the magnetic field increases, and at $\mu_0 H = 0.20$ T, the peak flattens out, providing qualitative evidence for the suppression of the cluster spin-glass phase. This observation is consistent with the Kondo breakdown scenario (a more detailed quantitative analysis is challenging; see fig. S15). Additionally, our dc susceptibility measurements exhibit a similar trend, where the FC and ZFC curves completely overlap at $\mu_0 H = 0.13$ T (Fig. S16).

**Conclusion and outlook**

We now attempt to combine the results of our measurements into a temperature-field phase diagram, as shown in Fig. 4. A fan-shaped quantum critical regime, marked by the logarithmic temperature dependence of the electronic specific heat coefficient, is observed between the cluster spin-glass and heavy-fermion liquid phases. The phase boundaries are defined by the temperature scale $T_{cs}$. Below $T_{cs}$, the heavy-fermion liquid ground state forms at higher magnetic fields ($\mu_0 H > \mu_0 H_c$), indicated by the formation of Kondo spin-singlets with large Fermi surfaces. At lower field ($\mu_0 H < \mu_0 H_c$), RKKY fluctuations disrupt the Kondo coherent composites, while strong magnetic frustration, originating from the random distribution of magnetic atoms, prevents LRMO, resulting in a cluster spin-glass state.

However, several unresolved issues remain regarding TiFe$_{0.7}$Cu$_{0.4}$Sb. Notably, charge fluctuations have recently been recognized as crucial to Kondo breakdown QPTs (*40*). Given the strong disorder in our system, these fluctuations may lead to inherently inhomogeneous states near the Kondo breakdown QPT. Furthermore, it is possible that the observed quantum critical behaviours occur in isolated "rare regions" of the sample, a phenomenon known as the quantum Griffiths effect (*29, 41*). This effect, which often arises in disordered systems, could explain the two-stage spin-glass freezing and the small values of measured quantities. Indeed, the broad peak maximum in specific heat (Fig. 2B) and magnetic Grüneisen parameter (Fig. 2E), and discrepancies between temperature and field scales across different experiments (Fig. 4) suggest that the QPT may be a "smeared" transition, indicating potential Griffiths regions in our samples. To elucidate these points, more detailed experimental studies, such as the synthesis of single crystals, and theoretical characterizations of the various phases are required.

Nonetheless, our conservative interpretation suggests that a magnetic field-induced QCP can develop within a spin-glass phase without associated symmetry breaking, filling a gap in our understanding of quantum criticality (fig. S19). Due to its unique lattice structure, $TiFe_xCu_{2x-1}Sb$ hosts randomly distributed magnetic moments (in clusters) coupled antiferromagnetically. This characteristic makes the material an intriguing platform for exploring quantum many-body physics, as posited by the Sachdev-Ye-Kitaev (SYK) model (*42*). The SYK model provides an exactly solvable framework incorporating randomly distributed coupling constants, $J_{ij}$. Interest in the SYK model spans high-energy and condensed matter physics, from Hawking's quantum information paradox in black holes to the strange metals observed in heavy-fermion materials. Remarkably, $TiFe_{0.7}Cu_{0.4}Sb$ exhibits a perfect linear temperature dependence in resistivity at temperatures well above room temperature of 300 K (fig. S4). This "bad metal" behaviour is also one of the predictions of the SYK model.

**Acknowledgments:** This work was supported by the National Key Research and Development Program of China (No. 2019YFA0704901) and National Natural Science Foundation of China (No. 92163212). W.Z. acknowledges the support from the Guangdong Innovation Research Team Project (No. 2017ZT07C062), Guangdong Major Talent project (No. 2019CX01C237), Shenzhen Municipal Key-Lab program (No. ZDSYS20190902092905285), Guangdong Provincial Key-Lab program (No. 2019B030301001). W.L. acknowledges the support from Shenzhen Key Program for Long-Term Academic Support Plan (No. 20200925164021002), Shenzhen Innovation Program for Distinguished Young Scholars (No. RCJC20210706091949018), and the support of the NSFC program for Distinguished Young Scholars (T2425012). Z.Y. acknowledges the support from National Natural Science Foundation of China for Distinguished Young Scholars (No. 52222102). Computing resources were supported by the Center for Computational Science and Engineering at the Southern University of Science and Technology. We thank Dr. Hanjie Guo from Songshan Lake Materials Lab in Dongguan for the assistance of ac susceptibility measurements.

**Author contributions:** W.Z. led and supervised the whole project. W.Z. and F.Z. conceived the idea, and with T.F., designed the experiments. The crystals were grown by T.F. and W.L., who also performed measurements of the electrical resistivity, specific heat, and dc magnetic susceptibility under various magnetic fields. M.H. performed electron backscatter diffraction experiments. X.Y. and Z.Y. performed scanning transmission electron microscopy measurements. Z.Z. and P.S. performed low-temperature dc magnetic susceptibility and magnetocaloric effect measurements. Y.R. and W.Z. performed *ab initio* calculations. F.Z. and W.Z. analyzed the raw experimental data, performed theoretical analysis and interpretations. L.Z., L.W., H.H. provided assistance in the experiments. Y.R. and B.W. provided assistance in the theories. F.Z. wrote the manuscript.

**Competing interests:** Authors declare that they have no competing interests.

**Data and materials availability:** The data that support the findings of this study are available from the corresponding author upon reasonable request.

**Corresponding author:** Correspondence and requests for materials should be addressed to yuzyemlab@fzu.edu.cn; liuws@sustech.edu.cn; zhangwq@sustech.edu.cn

**Supplementary Materials**
Materials and Methods, and Supplementary Text. figs. S1 to S19, References (43-59)

**Main Figure Legends**

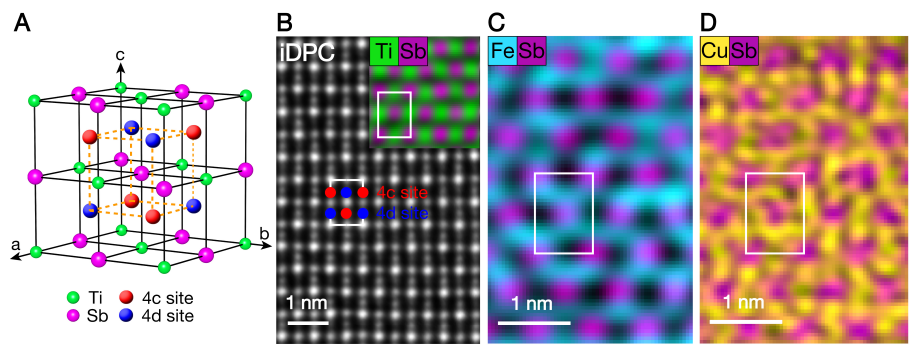

**Fig. 1. Crystal structure of TiFe$_{0.7}$Cu$_{0.4}$Sb**. **A,** Crystal structure of off-stoichiometry TiFe$_{0.7}$Cu$_{0.4}$Sb. Ti and Sb atoms form two interpenetrating face-centered cubic lattices, while Fe and Cu atoms partially and randomly occupy the 4*c* (red) and 4*d* (blue) sites within the structure. **B,** Integrated differential phase contrast (iDPC) image along the [110] direction. The white rectangle outlines a conventional unit cell, red and blue circles mark the positions of the 4*c* and 4*d* sites, respectively. Inset, energy dispersive spectroscopy (EDS) maps for Ti and Sb atoms, illustrating their periodic distribution and regular structural order. **C**, EDS maps for Fe atoms, and **D**, for Cu atoms, both highlighting their random distribution and indicating chemical disorder when occupying the 4*c* and 4*d* sites.

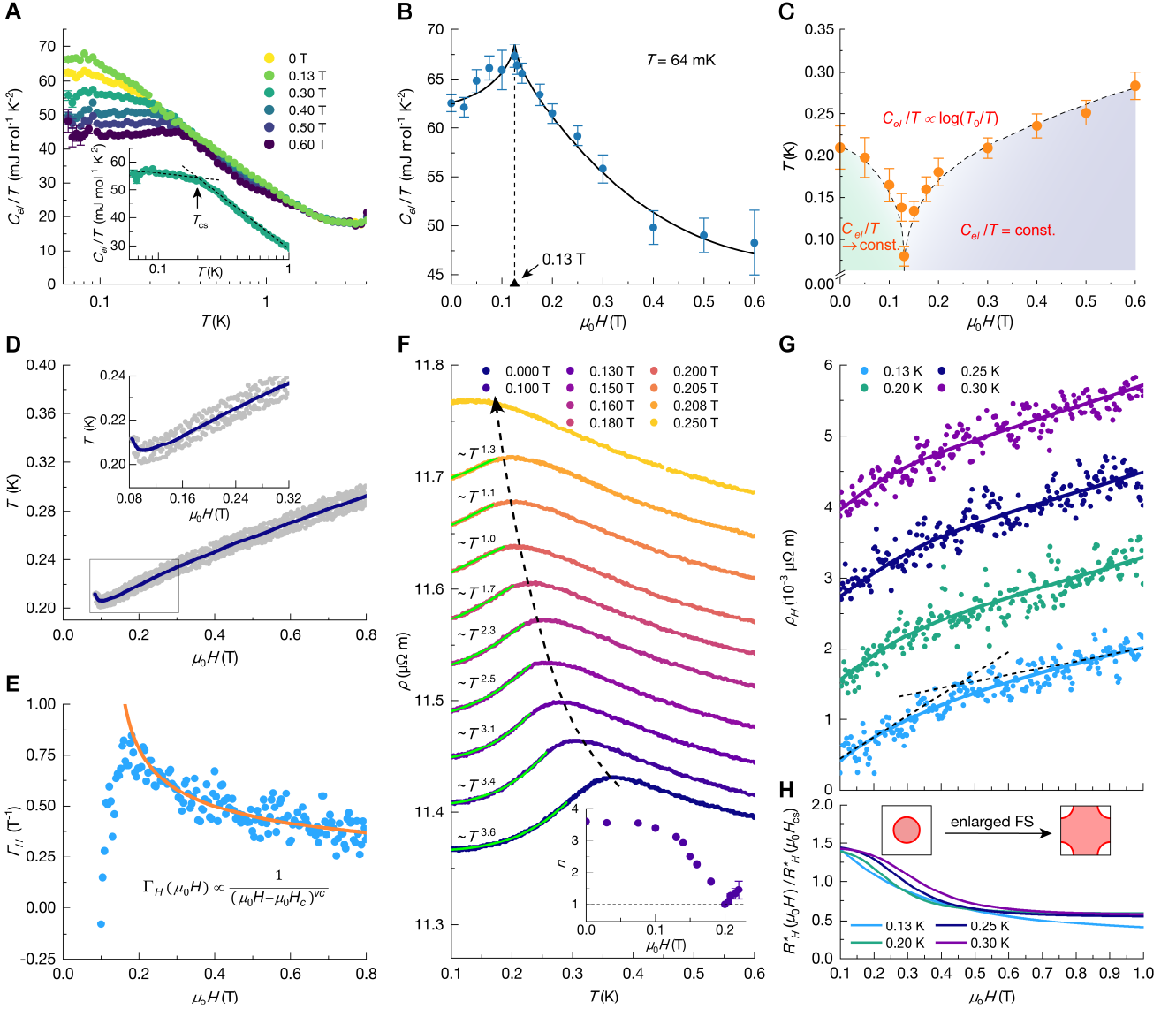

**Fig. 2. Thermodynamical and transport measurements of TiFe$_{0.7}$Cu$_{0.4}$Sb. A**, Electronic specific heat coefficient, $C_{el}(T)/T$, under various magnetic fields. Errors at low temperatures are due to nuclear Schottky contributions to specific heat (see Methods). Inset: $C_{el}(T)/T$ at $\mu_0 H = 0.30$ T. The arrow indicates the temperature scale $T_{cs}$ (see main text). **B**, $C_{el}(T)/T$ measured at $T = 64$ mK, which shows a pronounced peak near $\mu_0 H \approx 0.13$ T. The black solid line is a guide to the eye. **C**, The temperature scale $T_{cs}$ at various magnetic fields. Different regimes demonstrate different scaling behaviours of $C_{el}(T)/T$, as highlighted in red text. **D**, Magnetocaloric effect measured with varying magnetic fields down to $\mu_0 H = 0.08$ T. Below this field, the data becomes unreliable due to noise from magnetic flux transitions. Inset shows zoom in plot between $\mu_0 H = 0.08$ T and 0.32 T. **E**, Magnetic Grüneisen parameter, $\Gamma_H = dT/Td(\mu_0 H)$, as a function of the magnetic field $\mu_0 H$. The sold line is the best fit of data, with $\mu_0 H_c = 0.13$ T (see main text). **F**, Resistivity $\rho(T)$ under various magnetic fields. Data points are vertically shifted for better comparison. The low-temperature behaviour of resistivity is fitted to $\rho(T) \propto T^n$, with the exponent $n$ plotted in the inset. The dashed black curve with an arrow indicates the shift in the maximum of $\rho(T)$. **G**, Magnetic field dependence

of the Hall resistivity, $\rho_H(\mu_0 H)$, at different temperatures. Data points are vertically shifted for better comparison. Solid lines show the best fit, $\int R_H^*(\mu_0 H)dH$, to the data. Black dashed lines are guide to the eye. **H**, Effective Hall coefficient, $R_H^*(\mu_0 H)$, at various temperatures. As $\mu_0 H$ increases, $R_H^*(\mu_0 H)$ decreases from large to small values, indicating a transition from a small to a larger Fermi surface (FS) volume within the single-carrier model.

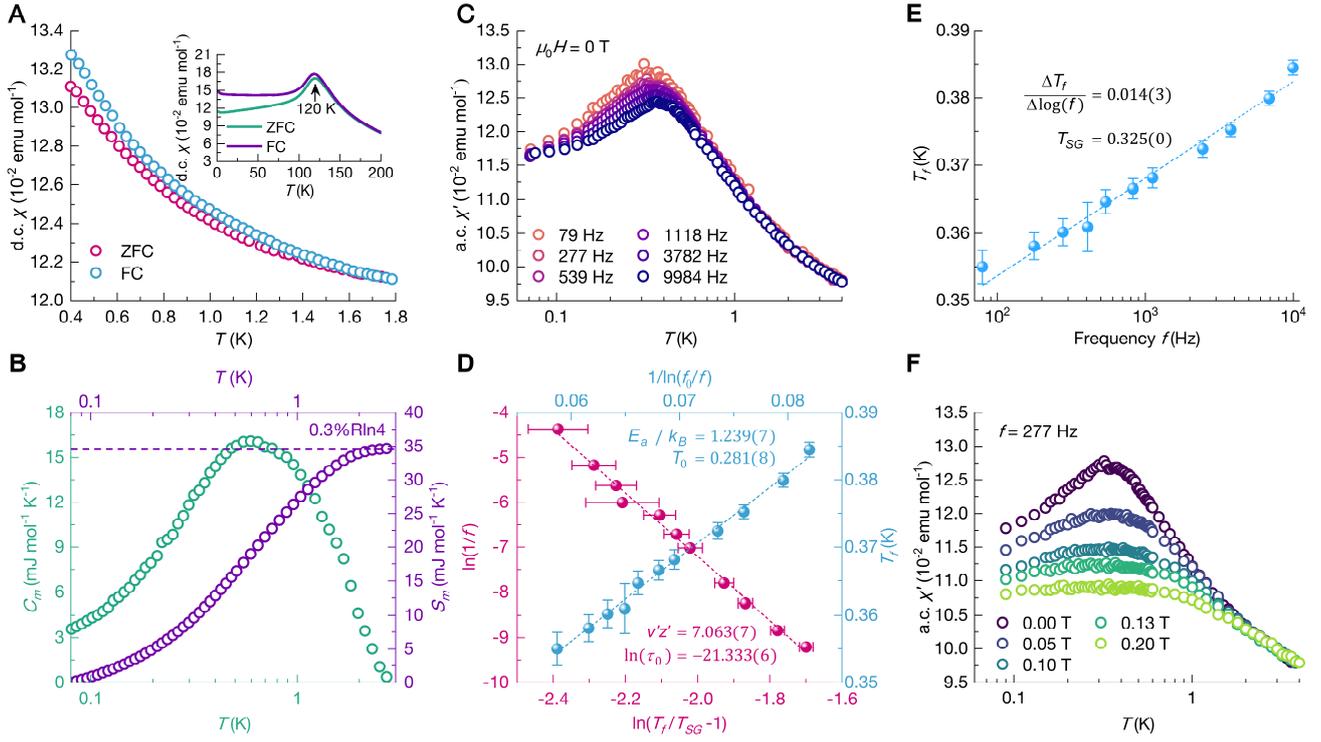

**Fig. 3. Magnetic measurements of TiFe$_{0.7}$Cu$_{0.4}$Sb. A**, dc magnetic susceptibility, $\chi(T)$, measured from 1.8 K down to 0.4 K. The inset shows $\chi(T)$ above 2 K, which has a peak at $T = 120$ K. **B**, Zero-field magnetic specific heat coefficient, $C_m(T)$ (see Methods), and the corresponding magnetic entropy, $S_m$. The calculated $S_m$ is approximately 4.3% of the expected value. **C**, Temperature and frequency dependence of the real part of the ac susceptibility, $\chi'(T)$, measured at zero magnetic field. **D**, The red data represents the linear fit of the freezing temperature as $\ln(T_f/T_{SG} - 1)$ versus $\ln(1/f)$. The intercept and slope provide values for $\ln(\tau_0)$ and $v'z'$, respectively. The blue data is the linear fit of the freezing temperature, $T_f$, versus frequency as $1/\ln(f_0/f)$. The intercept and slope yield values for $T_0$ and $E_a/k_B$, respectively (see main text for details). **E**, Frequency dependence of the freezing temperature, $T_f$, extracted from $\chi'(T)$ by fitting peaks using a Lorentzian function. The dashed line represents the best linear fit to the data points, with its intercept and slope providing values for $T_{SG}$ and $\Delta T_f/\Delta\log(f)$, respectively. **F**, Magnetic field dependence of $\chi'(T)$. With increasing $\mu_0 H$, the peak is gradually suppressed, indicating the weakening of the cluster spin-glass state.

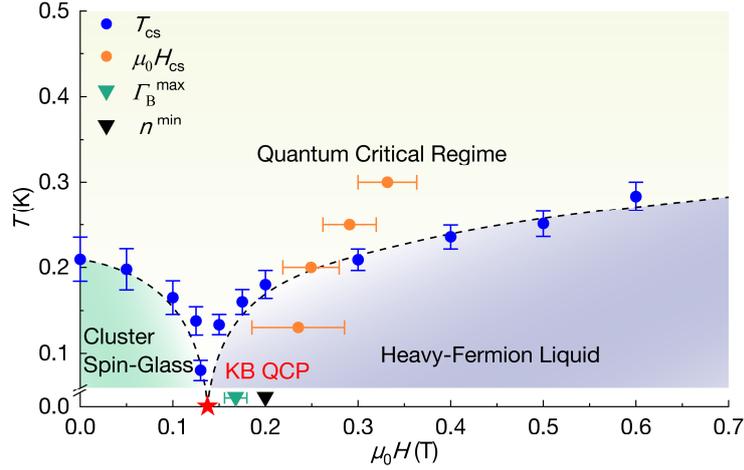

**Fig. 4. Derived phase diagram of TiFe$_{0.7}$Cu$_{0.4}$Sb.** Blue circles represent the temperature scale $T_{cs}$, derived from specific heat measurements, while orange circles indicate the magnetic field scale $\mu_0 H_{cs}$, obtained from Hall measurements. The green arrow points the magnetic field position where the magnetic Grüneisen parameter reaches its maximum, while the black arrow points the magnetic field position where the resistivity exponent $n$ approaches 1. The red pentagon indicates the proposed position of the QCP. Black dashed lines serve as visual guides. Crossing the Kondo breakdown (KB) QCP leads to a QPT from a cluster spin-glass to a heavy-fermion liquid as $T$ approaches zero.

Supplementary Materials for

# Emerging quantum critical phase in a cluster spin-glass

Fang Zhang[#], Tao Feng[#], Yurong Ruan, Xiaoyuan Ye, Bing Wen, Liang Zhou,
Minglin He, Zhaotong Zhuang, Liusuo Wu, Hongtao He, Peijie Sun,
Zhiyang Yu[*], Weishu Liu[*] & Wenqing Zhang[*]

[#]*These authors contributed equally to this work.*
*Corresponding authors: yuzyemlab@fzu.edu.cn, liuws@sustech.edu.cn; zhangwq@sustech.edu.cn*

**Materials and Methods**
1. Materials Preparation
2. Structure characterization
3. Thermodynamical, transport, and magnetic measurements
4. Ab initio calculations
5. Substruction of the nuclear Schottky contribution to the specific heat
6. Substruction of the phonon contribution to the specific heat
7. Substruction of the magnetic Schottky anomaly to the specific heat
8. Determination of scale $T_{cs}$ in electronic specific heat
9. Extracting the Hall resistivity $\rho_H$ from the raw data

**Supplementary Text**
10. Effective Hamiltonian of TiFe$_x$Cu$_{2x-1}$Sb
11. Fitting of the magnetic Grüneisen parameter
12. Estimation of the Kondo temperature $T_K$
13. Determination of the Kondo coherence temperature $T_{coh}$
14. ac and dc magnetic susceptibility at finite magnetic fields
15. Quantitative analysis of the crossover functions in Hall measurements
16. Kadowaki-Woods relation
17. Schematic of global phase diagram

**figure S1 to S19**
**Table S1**
**Reference [43-59]**

# Materials and Methods

1. **Materials Preparation**

The polycrystalline TiFe$_{0.7}$Cu$_{0.4}$Sb samples were synthesized through a combined approach of arc melting, mechanical alloying, and spark plasma sintering (SPS). The raw materials—Ti (99.99% pure rods), Fe (99.98% pure sheets), Cu (99.9% pure rods), and Sb (99.99% pure rods)—were weighed according to the nominal composition of TiFe$_{0.7}$Cu$_{0.4}$Sb. Firstly, the alloy ingot was prepared by arc melting on a water-cooled copper hearth under an Ar-protected atmosphere. The ingot was flipped over and re-melt for four times to ensure better homogeneity. A small amount of extra Sb was added to compensate the weight loss of Sb due to its high vapor pressure. Then the ingot was loaded into a stainless-steel ball milling jar in a glove box under an Ar atmosphere with an oxygen level of <1 ppm. After ball milling for 12 min in a SPEX 8000M mixer, the ball-milled powders were loaded into a graphite die with an inner diameter of 12.7 mm, in the glove box. The graphite die with the loading powder was immediately sintered at 750°C under a pressure of 50 MPa for 5 min via spark plasma sintering (SPS) (SPS-211Lx, Fuji Electronic Industrial Co. LTD).

2. **Structure characterization**

The phase purity of the product was measured by powder X-ray diffraction (XRD) on a Rigaku D/Max-2550 instrument (Cu K$\alpha$ radiation, $\lambda = 1.5418$ Å, 18 KW). The microstructures of the samples were examined by a high-resolution transmission electron microscopy (HRTEM) (JEM-F200, JEOL, Japan) and a probe Cs-corrected TEM (Themis ETEM, Thermo Fisher Scientific, USA). TEM specimens were prepared by mechanical slicing, polishing, and dimpling, followed by ion-milling. Energy-dispersive spectroscopy (EDS) was used to determine the distribution of elements at the nanoscale.

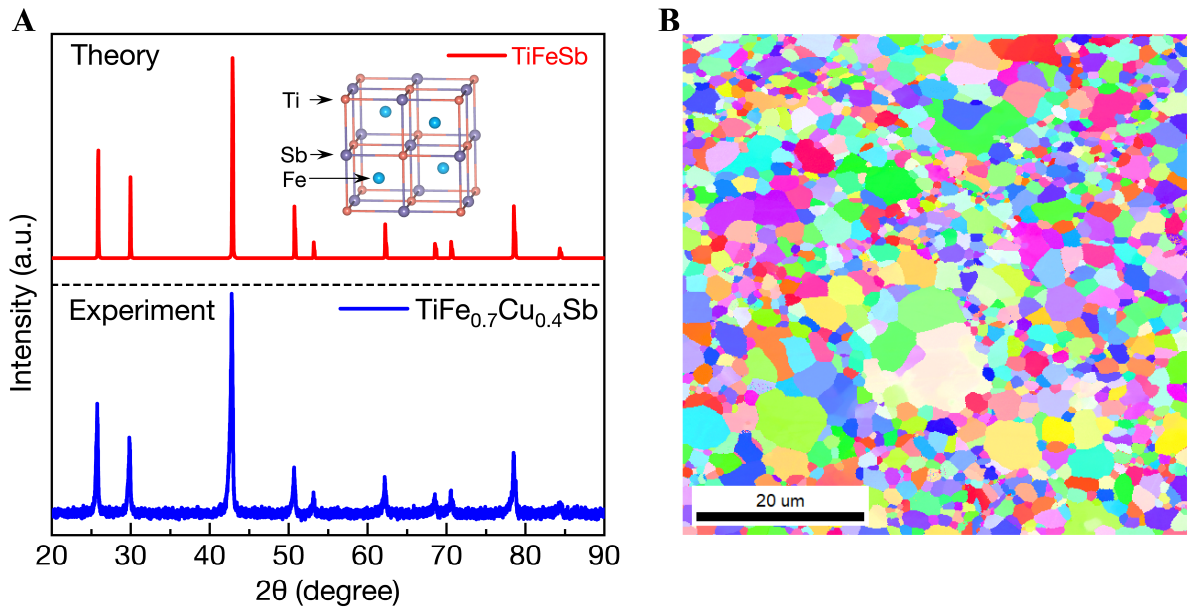

**fig. S1: XRD and EBSD patterns of TiFe$_{0.7}$Cu$_{0.4}$Sb. A,** The upper panel shows the theoretical X-ray diffraction (XRD) pattern of TiFeSb, a half-Heusler material where Fe occupies all 4c sites. The inset depicts the crystal structure. The lower panel displays the experimental XRD pattern of TiFe$_{0.7}$Cu$_{0.4}$Sb, confirming phase purity and the absence of impurities. **B,** The electron backscatter diffraction (EBSD) pattern of polycrystalline TiFe$_{0.7}$Cu$_{0.4}$Sb, highlighting the size and distribution of the grains.

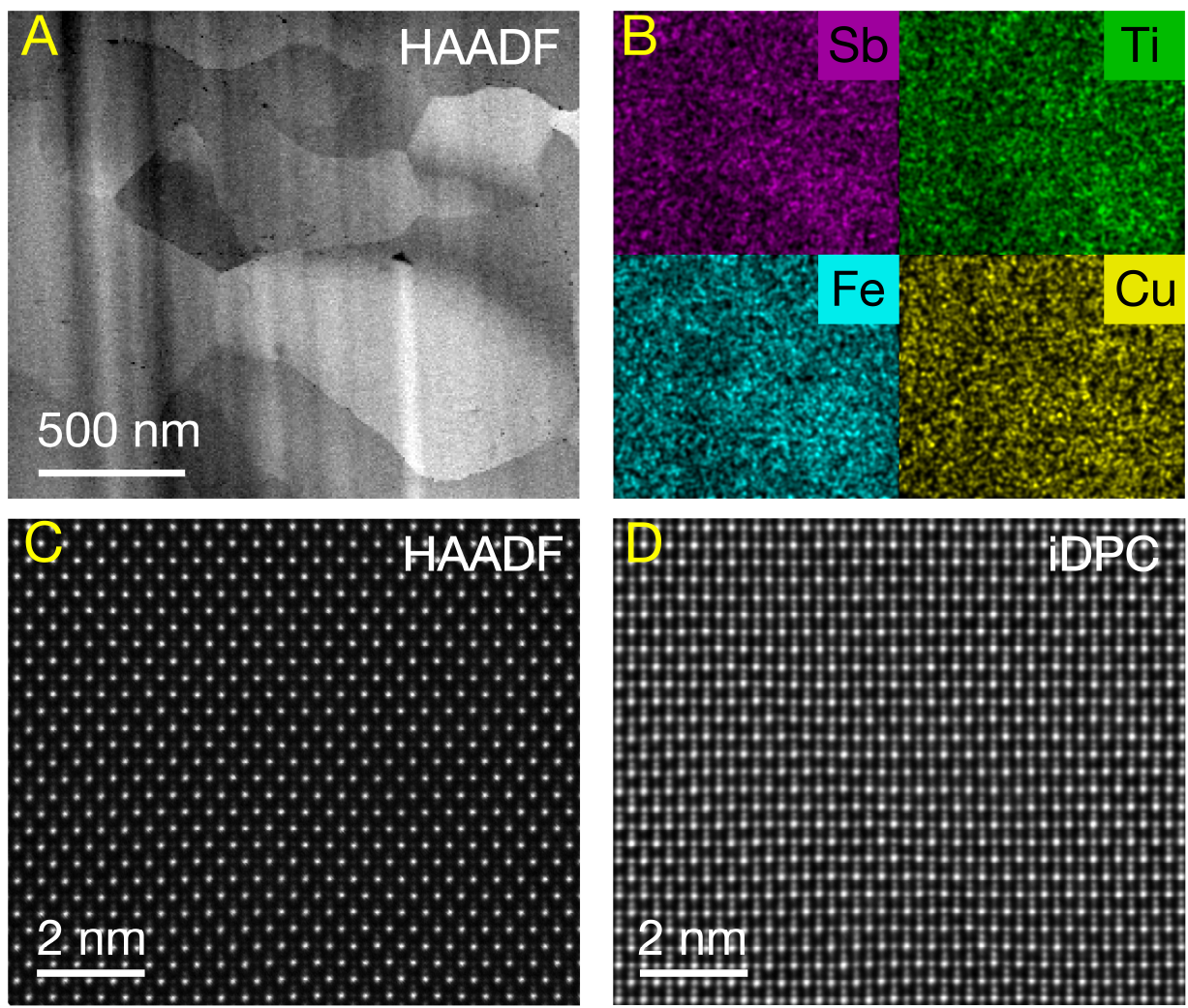

**fig. S2: Microstructures of TiFe$_{0.7}$Cu$_{0.4}$Sb. A**, High-angle annular dark-field scanning transmission electron microscopy (HAADF-STEM) image of TiFe$_{0.7}$Cu$_{0.4}$Sb. **B**, corresponding energy dispersive spectroscopy (EDS) maps showing the uniform distribution of Ti, Fe, Cu, and Sb elements within the sample. **C**, HAADF-STEM image along the [110] direction. **D**, integrated differential phase contrast (iDPC) image along the [110] direction.

## 3. Thermodynamical, transport, and magnetic measurements

The heat capacities were performed by Quantum Design Physical Property Measurement System (PPMS, 14T) with a dilution refrigerator (DR) insert in the temperature range of 0.05 K to 4 K at a set of fixed magnetic fields from 0 to 0.6 Tesla. The sample was mounted to the sample platform with a small amount of Apiezon grease and measured with the standard relaxation method.

The magnetocaloric effect (MCE) was evaluated by using a homemade quasi-adiabatic sample stage adapted for the Oxford $^3$He refrigerator. The sample stage consists of a 3×3 mm² sapphire plate suspended by thin nylon wires within a PEEK frame. A resistance thermometer (CX-1010 bare chip sensor, Lake Shore Cryotronics Inc.), fixed to the sapphire plate and connected by 25 $\mu$m diameter manganin wires, was used to monitor the temperature changes during MCE measurement.

A lock-in technique was applied to measure the resistivity, magnetoresistivity and Hall resistivity. The sample was loaded on the platform of DR insert of PPMS, with the temperature range between 0.05 K and 4 K. For resistivity and magnetoresistivity measurements, the sample was bonded using the four-contact configuration, and a relatively large current of 150 $\mu$A is used due to the small value of the resistivity. Because of the heating effects of the current, the minimum temperature we can reach is near 0.1 K instead of the designed 0.05 K (see fig. S3). For Hall effect measurement, the sample was bonded as standard Hall bar. The ac current was set with an amplitude of 100 $\mu$A and a frequency of 37 Hz. Due to the magnetic flux jump noise of magnets, the minimum magnetic fields $\mu_0 H$ we present is 0.1 T, as the temperature is not stable for smaller fields (see Sec. 9 for more details).

The low-temperature (0.4 K to 1.8 K) dc magnetic susceptibility was measured by using a vibrating sample magnetometer equipped with a SQUID sensor (SQUID-VSM, Quantum Design) in conjunction with a $^3$He insert (iHelium3). The high-temperature (above 2 K) dc magnetic susceptibility was measured by the VSM option of PPMS with an applied magnetic field of 0.01T. The ac magnetic susceptibility was performed on ACDR option of PPMS in the temperature range between 0.05 K and 4 K at a set of fixed frequency from 79 Hz to 9984 Hz. The AC Drive was set as 3 Oe.

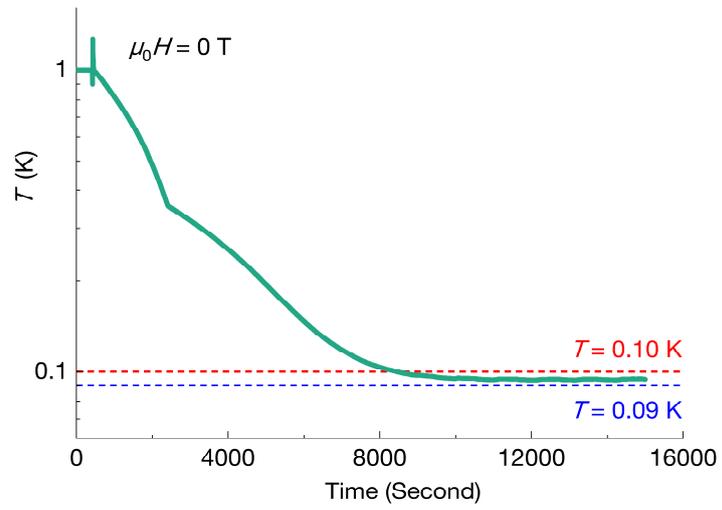

**fig. S3: Temperature record.** A typical temperature record during resistivity measurement. Due to heating from the applied current, the minimum achievable temperature is around 0.10 K.

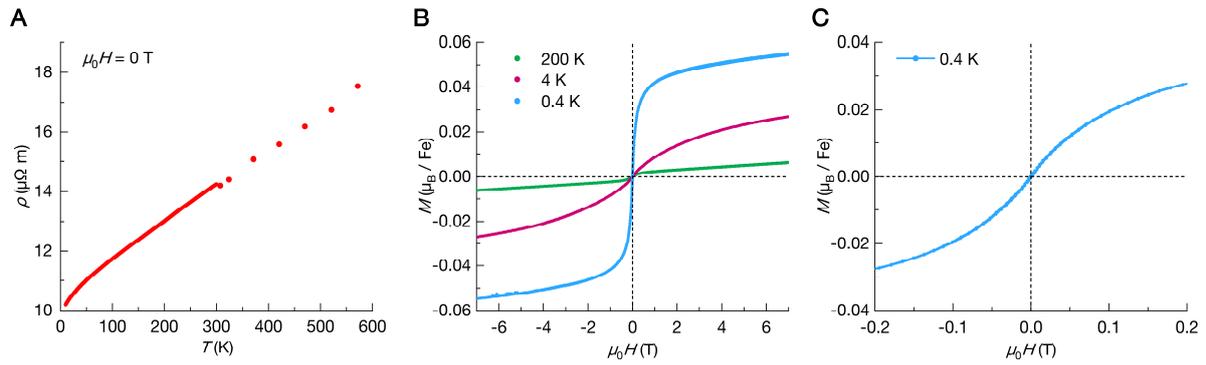

**fig. S4: High-temperature resistivity and field-dependent magnetization. A**, The resistivity from 2 K to 600 K at zero field shows metallic behaviour. A linear temperature dependence is observed from 100 K to 600 K, characterizing TiFe$_{0.7}$Cu$_{0.4}$Sb as a "bad metal". **B**, Magnetization at 0.4 K, 4 K, and 200 K. With increasing the magnetic fields, magnetization increases and saturates at approximately 0.05 $\mu_B$ per Fe atom at 0.4 K and 5.0 T. **C**, Magnetization loops at 0.4 K, which shows no hysteresis, ruling out the ferromagnetic ordering in the system.

### 4. Ab initio calculations

Density functional theory (DFT) calculations were performed using the Vienna Ab initio Simulation Package (VASP) *(43, 44)*. TiFe$_{0.7}$Cu$_{0.4}$Sb compound was calculated through the employment of a 3 × 3 × 3 supercell Ti$_{108}$Fe$_{76}$Cu$_{44}$Sb$_{108}$, where Fe and Cu atoms are randomly distributed along the 4c and 4d sites. The exchange-correlation functional adopted the Strongly Constrained and Appropriately Normed (SCAN) *(45)*. For the Brillouin zone integration, a gamma-centered 2 × 2 × 2 **k**-point mesh was selected, with a cutoff energy set at 400 eV. All atoms were fully relaxed until the residual forces acting on each atom diminished below 0.02 eV/Å. An energy convergence criterion of 10$^{-6}$ eV was set to guarantee the attainment of highly converged results.

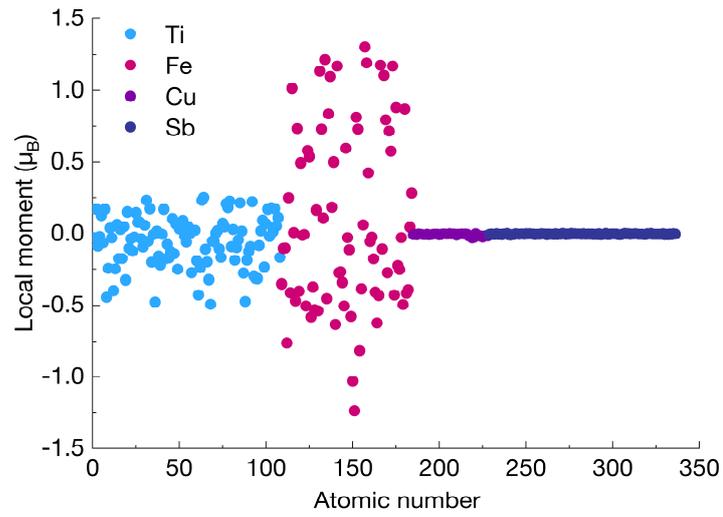

**fig. S5: Calculated distribution of magnetic moments in TiFe$_{0.7}$Cu$_{0.4}$Sb.** Ab initio calculations revealed that magnetic moments primarily originate from Fe atoms, with a minor contribution from Ti atoms. Due to strong disorder, Fe atoms experience varying crystal field environments, leading to a range of magnetic moments between 0 and 1.5 $\mu_B$.

## 5. Substruction of the nuclear Schottky contribution to the specific heat

As shown in fig. S6, the raw specific heat data $C_{tot}(T)$ without any correction for zero and finite fields exhibits a sharp increase at very low temperatures on cooling. This phenomenon may be caused by the spin-glass melting down, or a nuclear Schottky contribution. However, the real part of ac susceptibility $\chi'(T)$ smoothly decreases on cooling (Fig. 3C in main text), ruling out the spin-glass melting down for which would result in an increasing of the spin susceptibility. The nuclear Schottky anomaly is due to the nuclear level of Fe atoms. Even at zero magnetic field, the existence of strong lattice disorder distorts the tetrahedron crystal field of Fe, creating finite electrical field gradient that splits its nuclear levels.

Generally, the nuclear Schottky contribution to the specific heat $C_{nuc}(T)$ is

$$C_{nuc}(T) = Nk_B \frac{\alpha^2}{4I^2}\left[\frac{1}{\sinh^2\left(\frac{\alpha}{2I}\right)} - \frac{(2I+1)^2}{\sinh^2\left(\frac{(2I+1)\alpha}{2I}\right)}\right], \quad \alpha = \frac{A_{hf}\mu I}{\mu_B g_I T} \quad (S1)$$

where $N$, $k_B$ and $\mu_B$ are Avogadro's number, Boltzmann's constant, and Bohr's magneton. $A_{hf}$, $\mu$, $I$ and $g_I$ are hyperfine coupling, Fe-ion magnetic moment, nuclear spin, and Lande's $g$-factor. Since these parameters are mostly unknown, direct estimation of the $C_{nuc}(T)$ based on eq. S1 is unavailable.

We estimate $C_{nuc}(T)$ following the method in ref *(46)*. Firstly, to a good approximation, eq. S1 can be expressed as *(47)*

$$C_{nuc}(T) \approx \frac{A}{T^2} \quad (S2)$$

Then, we estimated the coefficient $A$ by linear-fitting the low-temperature $C_{tot}(T)$ with $C_{tot}(T)/T \sim A/T^3$ at $T \leq 0.1$ K regions. The fitted $A$ is summarized in table S1 with standard error. All fits have a confidence interval greater than 99.7%.

The Sommerfeld coefficient $\gamma(T)$ at $T \to 0$ reported in Fig. 2B is obtained by $\gamma(T) = [C_{tot}(T) - C_{nuc}(T)]/T$ calculated at $T = 64$ mK, whereas the large error bar comes from the fittings of coefficient $A$. At this low temperature, the phonon and magnetic Schottky contribution to the specific heat is negligible (as will be described next).

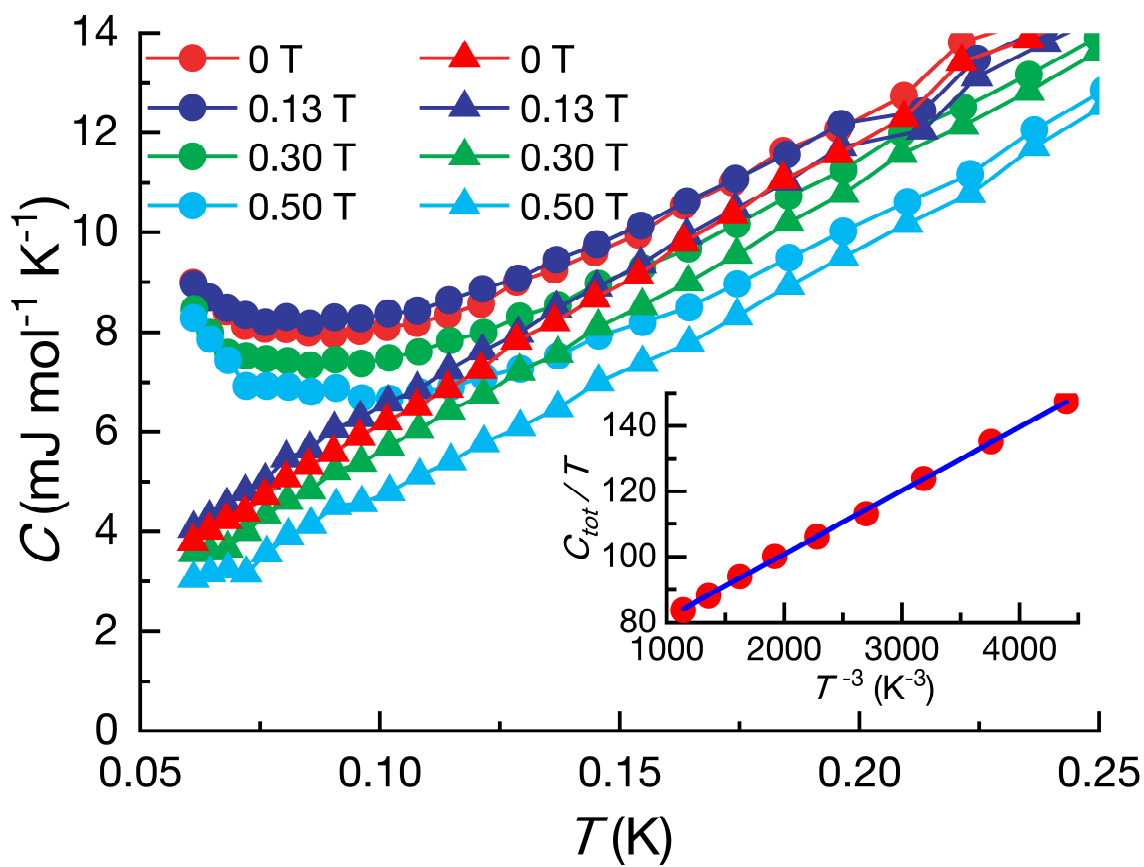

**fig. S6: Substruction of the nuclear contribution $C_{nuc}$ to specific heat.** The raw specific heat data (circles), $C_{tot}$, shows a sharp increase at low temperatures on cooling, potentially due to a nuclear Schottky contribution. After subtracting this contribution, the corrected specific heat $C_{tot} - C_{nuc}$ is plotted (triangulars). The inset shows $C_{tot}/T$ versus $1/T^3$ in the $T \leq 0.1$ K region, with a linear fit to the data.

## 6. Substruction of the phonon contribution to the specific heat

In normal metals, the low-temperature specific heat is contributed by conduction electron and phonon as $C(T) = \alpha T + \beta T^3$ (48). If plotted as $C(T)/T \sim T^2$, the result should be a line. However, as shown in fig. S7, the specific heat $[C_{tot}(T) - C_{nuc}(T)]/T$ in TiFe$_{0.7}$Cu$_{0.4}$Sb, where the nuclear Schottky contribution has been subtracted, is a line with respect to $T^2$ at high-temperature regions but deviates extensively at low temperature, which is a typical characteristic of heavy-fermion materials. The low-temperature anomaly is attributed to magnetism, namely the interaction between conduction electrons and local moments, as $C_m(T)$.

The phonon contribution to the specific heat $C_{pho}(T)$ is obtained by linear-fitting $[C_{tot}(T) - C_{nuc}(T)]/T$ with respect to $T^2$ at $T > 2.5$ K at $\mu_0 H = 0$ T, and the fitted value is $\beta = 0.00105 \pm 0.00006$ J mole-Fe$^{-1}$ K$^{-4}$. Since magnetic fields do not affect the crystal structure and hence $C_{pho}$, the same $\beta$ was employed at finite fields.

The intercept obtained in above fitting procedure is the contribution of normal conduction electrons, i.e., $\alpha T$ with $\alpha = 0.01850 \pm 0.00067$ J mole-Fe$^{-1}$ K$^{-2}$.

The magnetic contribution to specific heat $C_m$ as plotted in Fig. 3B is obtained by $C_m(T) = C_{tot}(T) - C_{nuc}(T) - C_{pho}(T) - \alpha T$.

In a free electron model, $C_{el} = \frac{3}{2} R \left[\frac{1}{3} \pi^2 \frac{T}{T_F}\right] = \gamma_0 T$, and $\frac{m^*}{m} = \frac{\gamma}{\gamma_0}$. In TiFe$_{0.7}$Cu$_{0.4}$Sb, $\gamma \approx 62$ mJ mole-Fe$^{-1}$ K$^{-2}$ at $T \approx 0.08$ K. If we approximate the Fermi temperature to $10^5$ K, which is a typical value for metals, the low temperature effective mass of TiFe$_{0.7}$Cu$_{0.4}$Sb is estimated to be 151 times to that of free electrons, suggesting TiFe$_{0.7}$Cu$_{0.4}$Sb is a heavy-fermion metal.

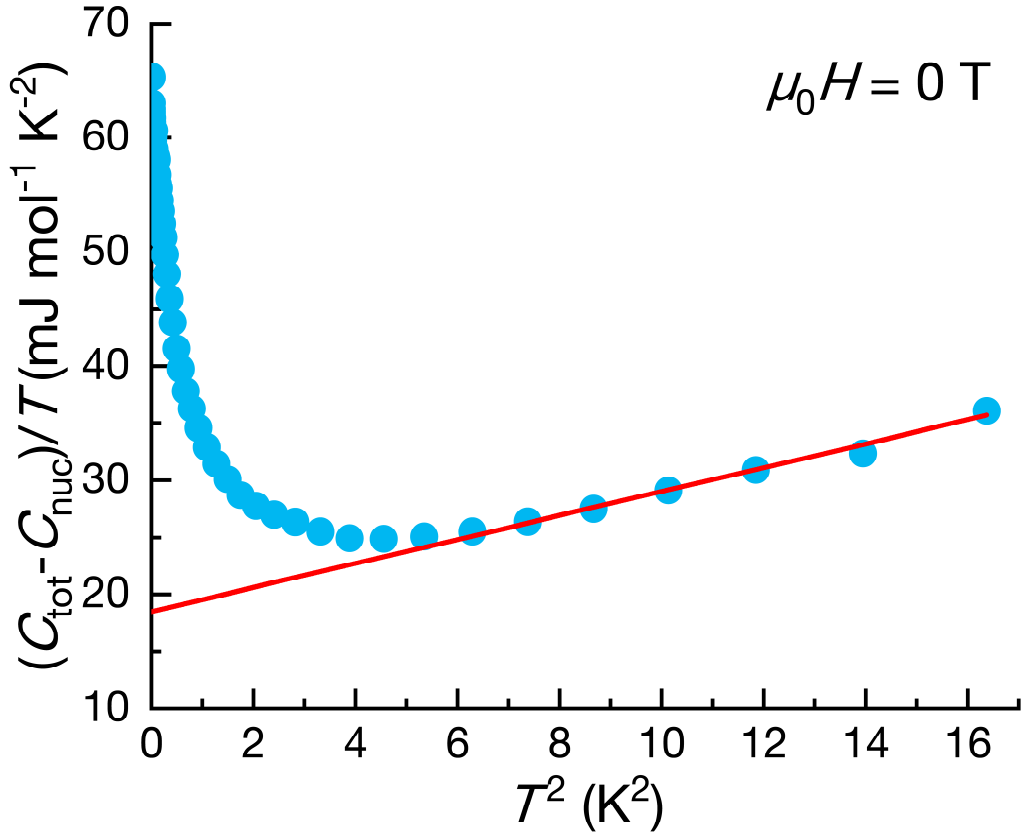

**fig. S7: Substruction of the phonon contribution $C_{pho}$ to specific heat.** The circles represent the specific heat after the subtraction of the nuclear contribution, $C_{tot} - C_{nuc}$. At higher temperatures, the relation $C_{tot} - C_{nuc} = \alpha T + \beta T^3$ holds, with the red line indicating the best fit above 2.5 K ($\alpha = 0.01850 \pm 0.00067$ J mole-Fe$^{-1}$ K$^{-2}$, $\beta = 0.00105 \pm 0.00006$ J mole-Fe$^{-1}$ K$^{-4}$). The low-temperature deviation signifies the presence of magnetic specific heat, $C_m$, due to spin fluctuations. $C_m$ can be obtained by $C_m = C_{tot} - C_{nuc} - \alpha T - \beta T^3$, where $C_{pho} = \beta T^3$.

## 7. Substruction of the magnetic Schottky anomaly to the specific heat

In Fig. 2A, we plotted the electronic specific heat coefficient $C_{el}(T)$. In addition to the nuclear Schottky contribution $C_{nuc}(T)$ and the phonon contribution $C_{pho}(T)$ to the total specific heat $C_{tot}(T)$, there is also a magnetic Schottky anomaly $C_{sch}(T)$ arising from the Kondo effect, which is related to single-ion impurity scattering. Specifically, $C_{sch}(T)$ can be calculated as follows:

$$C_{sch}(T) = n \frac{2N\Delta^2}{k_B T^2} \frac{e^{\Delta/k_B T}}{(1 + 2e^{\Delta/k_B T})^2} \quad (S3)$$

where $N$ and $k_B$ are Avogadro's number and Boltzmann's constant, respectively. Here, $\Delta$ is the energy level splitting, and $n$ counts how many local moments (per mole-Fe) contributes to $C_{sch}$.

In TiFe$_{0.7}$Cu$_{0.4}$Sb, due to the presence of strong disorder, the local moments (mostly originating from Fe atoms) vary from 0 to 1.5 $\mu_B$. In the main text, we take an average value of 0.05 $\mu_B$ as obtained from magnetization measurements. To estimate the parameters $n$ and $\Delta$ (based on a two-energy level system assumption), we take a more simplified approach, assuming that the local moments can only take values of either 0 or 1.5 $\mu_B$. Only Fe atoms with a magnetic moment of 1.5 $\mu_B$ contribute to $C_{sch}$, while those with zero magnetic moment do not. In this physical model, each Fe atom either contributes a finite and same value to $C_{sch}(T)$ or none at all, which simplifies the estimation.

We first estimate $\Delta$. According to this model, the broad peak $C_m(T)$ at $T^+ = 0.60$ K (see Fig. 3B in the main text) is attributed to the Kondo effect of these magnetic Fe atoms with a moment of 1.5 $\mu_B$. Therefore, we estimate the energy splitting $\Delta$ by finding $\frac{\partial C_{sch}}{\partial T}\big|_{T=T^+} = 0$, leading to:

$$2e^{\Delta/k_B T^+} = \frac{\Delta/k_B T^+ + 2}{\Delta/k_B T^+ - 2} \quad (S4)$$

For a rough estimate of Eq. S4, $\Delta \geq 2k_B T^+$. The actual value of $\Delta$ is calculated numerically and summarized in Table S1. Increasing magnetic fields gradually enhance the Kondo coupling strength $J_K$ due to the increasing of spin fluctuation, resulting in a monotonic increase in $\Delta$.

The value of $n$ is roughly estimated by calculating the magnetic entropy. Specifically, for a spin-$S$ particle, the theoretical magnetic entropy is $S_m^0 = R\ln(2S + 1)$ if all local moments are considered. At $\mu_0 H = 0$ T, the calculated magnetic entropy $S_m$ is 34.6 mJ mol$^{-1}$ K$^{-1}$, as elaborated in the main text. Here, for $S = 3/2$, corresponding to a magnetic moment of 1.5 $\mu_B$, the theoretical magnetic entropy is 11520 mJ mol$^{-1}$ K$^{-1}$. Thus, for a rough estimate,

$$n \sim \frac{S_m}{S_m^0} = 0.3\% \quad (S5)$$

This indicates that, if all magnetic Fe ions had moments of 1.5 $\mu_B$, only approximately 0.3% of the Fe atoms contribute to the magnetic moments at low temperatures, while the remainder are nonmagnetic. The reason for this small fraction of magnetic Fe is due to the following effects: (1) Strong disorder in the system results in different crystal field environments for each Fe ion. Some Fe ions experience a crystal field environment similar to that found in half-Heusler (non-magnetic) or full-Heusler (weakly magnetic) compounds, resulting in very small magnetic moments; (2) As discussed in the main text, spin-glass freezing occurs at $T_f' = 120$ K. Consequently, at low temperatures (below 1 K), most spins are frozen and do not participate in the quantum critical process.

The actual value of $n$ used under different magnetic fields are summarized in table S1. Considering the uncertainty in obtaining $C_m$ by subtracting other contributions to specific heat and the potential underestimation of $S_m$ due to a finite temperature range, the values of $n$ used are very close to 0.3% (ranges from 0.28% to 0.45% with increasing fields) but not exactly equals to it. The actual values of $n$ were determined by subtracting the peak at $T = T^+$ in $C_{el}(T)$, where $C_{el}(T) = C_{tot}(T) - C_{nuc}(T) - C_{pho}(T) - C_{sch}(T)$. $C_{el}(T)$ at different magnetic fields is plotted in fig. S8.

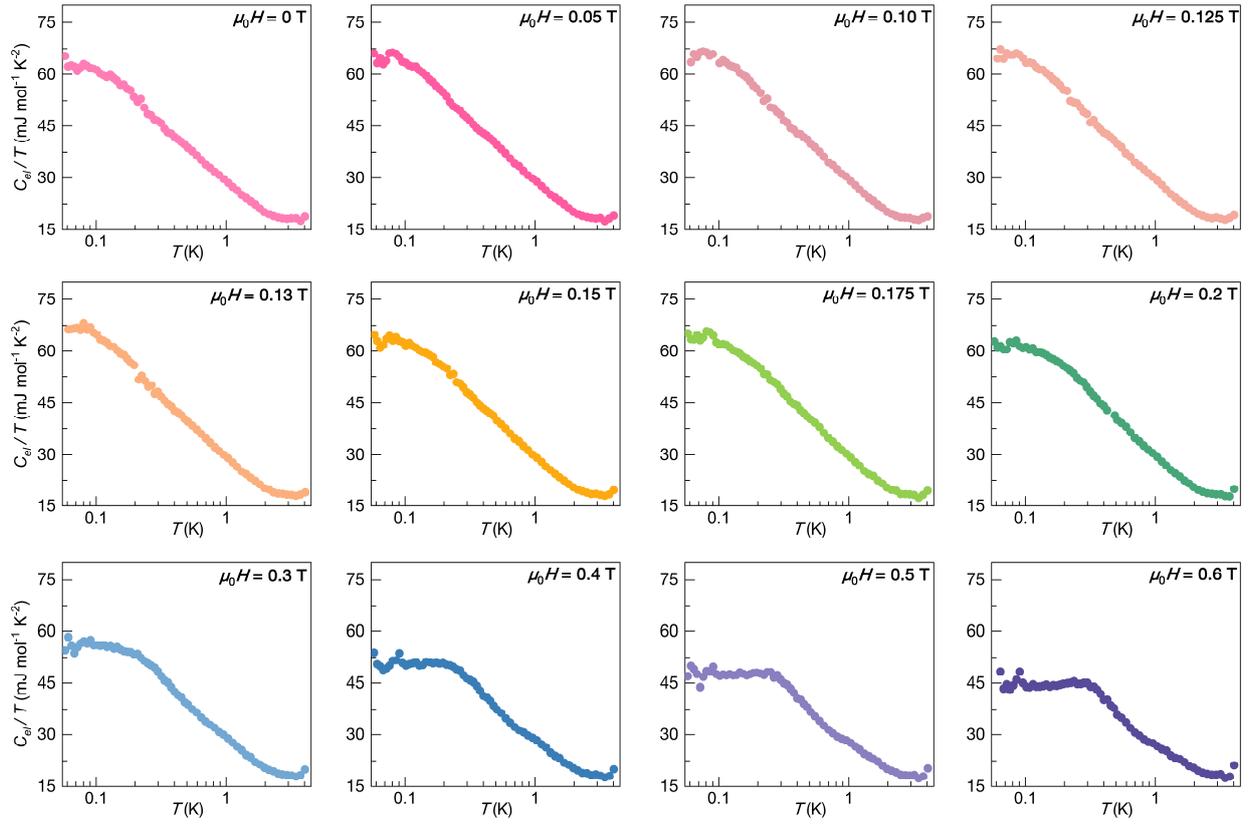

**fig. S8: Electronic specific heat coefficient $C_{el}$.** The electronic specific heat coefficient is determined as $C_{el} = C_{tot} - C_{nuc} - C_{pho} - C_{sch}$, plotted as $C_{el}/T$ for different magnetic fields.

## 8. Determination of scale $T_{cs}$ in electronic specific heat

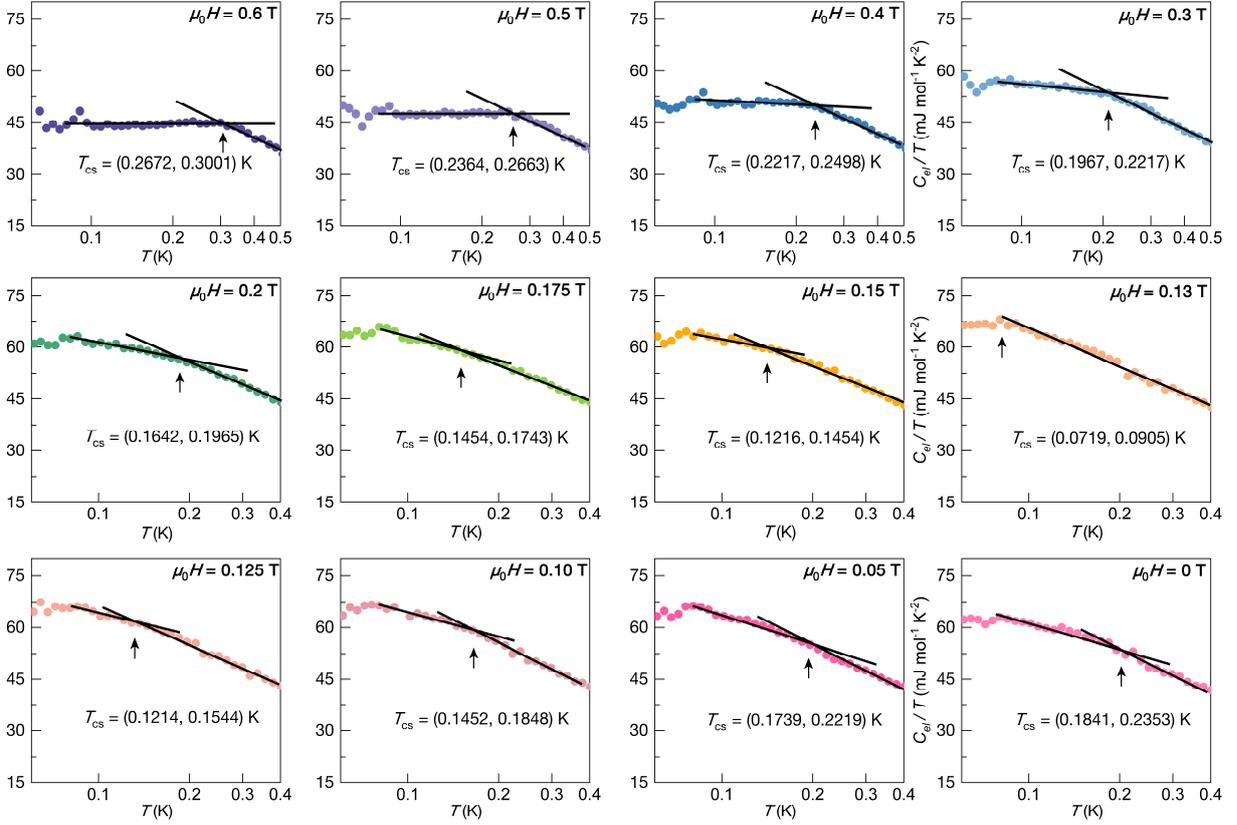

**fig. S9: Determination of scale $T_{cs}$.** Above and below $T_{cs}$, $C_{el}/T$ can be fitted by the logarithmic form $C_{el}/T = C_{el}^0 + A\log(T)$, but with different values of $A$. Below $T_{cs}$, the absolute value of $|A|$ is smaller. To determine $T_{cs}$, we try several trial temperatures $T_{cs}^0$ and perform fits to search for the temperature that yields high adjusted R-squared values in both the high- and low-temperature regions. This temperature is then determined to be $T_{cs}$. Due to data scattering below 0.08 K (partly due to uncertainties in the nuclear Schottky contribution, $C_{nuc}$), we only consider the temperature regime $T \geq 0.08$ K during the fitting process. Additionally, because of the limited number of data points, there may be cases where only one trial temperature provides a satisfactory fit. In such situations, we define the uncertainty in $T_{cs}$ by the temperature step size in our measurements.

## 9. Extracting the Hall resistivity $\rho_H$ from the raw data

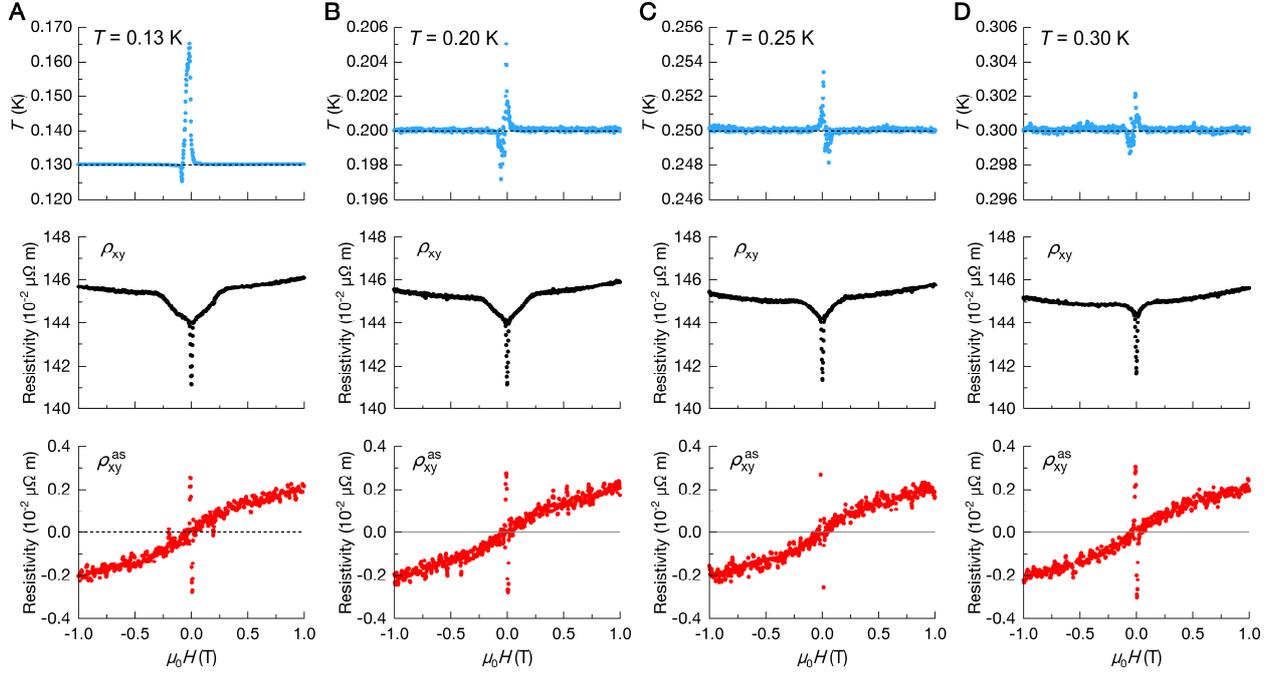

**fig. S10: Anti-symmetrization of Hall resistivity.** The raw Hall resistivity $\rho_{xy}$ contains contributions from magnetoresistivity. To obtain the intrinsic Hall resistivity, we calculate the anti-symmetrized Hall resistivity as $\rho_{xy}^{as}(\mu_0 H) = \left[\left(\rho_{xy}(\mu_0 H) - \rho_{xy}(-\mu_0 H)\right)\right]/2$, where $\rho_{xy}(\mu_0 H)$ is the measured raw Hall resistivity, and $\rho_{xy}^{as}(\mu_0 H)$ is taken as the real Hall resistivity, denoted as $\rho_H$. The first row shows the recorded temperature during the measurements. When the magnetic field $\mu_0 H$ is flipped from negative to positive values, the temperature becomes unstable. The second row displays the raw Hall resistivity $\rho_{xy}$, which includes significant contributions from magnetoresistivity. The third row presents the anti-symmetrized resistivity $\rho_{xy}^{as}$, which we use as the real Hall resistivity, $\rho_H$. Due to temperature instability during magnetic field flips, in the main text we only consider the regime where the magnetic field $\mu_0 H > 0.1$ T.

# Supplementary Text

## 10. Effective Hamiltonian of TiFe$_x$Cu$_{2x-1}$Sb

Due to the complex structure of TiFe$_x$Cu$_{2x-1}$Sb, its effective Hamiltonian is composed of following terms

$$H = H_c + H_d + H_I \tag{S6}$$

where $H_c$ comes from conduction electrons, $H_d$ is from localized electrons hosting local moments, $H_I$ is the hybridization between conduction electron and localized electrons.

The conduction electrons in TiFe$_x$Cu$_{2x-1}$Sb are contributed by the partially itinerant 3$d$-electrons from Ti atoms, and the itinerant 5$p$-electrons from Sb atoms (copper is used to modify the electrons concentration).

$$H_c = \sum_{ij\sigma} t^1_{ij} a^\dagger_{i\sigma} a_{j\sigma} + \sum_{ij\sigma} t^2_{ij} b^\dagger_{i\sigma} b_{j\sigma} + \sum_{ij\sigma} t^0_{ij} a^\dagger_{i\sigma} b_{j\sigma} + h.c. \tag{S7}$$

where $a^\dagger_{i\sigma}$ and $b_{j\sigma}$ are the creation operator of 3$d$-electrons from Ti atoms and 5$p$-electrons from Sb atoms, respectively; $\sigma$ denotes the spin index, ↑ and ↓; $t^1_{ij}$, $t^2_{ij}$, $t^0_{ij}$ is the hopping amplitude between Ti-Ti sites, Sb-Sb sites and Ti-Sb sites. h.c. stands for Hermitian conjugate. Generally, Ti and Sb are different sites with different types of electrons; however, to our purpose here which only acquire their itinerant nature (the correlation effect of Ti $d$-electrons and the small magnetic moments are thus neglected), we may consider $d$-electrons from Ti and $p$-electrons from Sb as the "same type" $a^\dagger_{i\sigma} \approx b^\dagger_{i\sigma} = c^\dagger_{i\sigma}$ with different "orbital" flavors, $c^\dagger_{i\sigma} \rightarrow c^\dagger_{i\sigma\alpha}$ ($\alpha$ refers to flavor index). Eq. S7 can be rewritten as

$$H_c = \sum_{ij\sigma\alpha} t_{ij} c^\dagger_{i\sigma\alpha} c_{j\sigma\alpha} \tag{S8}$$

where $t_{ij}$ is the hopping amplitude between different lattice and basis sites.

The localized electrons which host local moments are mostly originated from Fe atoms. The three-fold degenerate $t_{2g}$ orbitals of 3$d$-electrons in Fe are full occupied, whereas the two-fold degenerate $e_g$ orbital is only occupied by one electron, and thus the local spins have a degeneracy $N = 4$. The Hamiltonian from Fe $d$-electrons is composed of

$$H_d = \sum_{i\sigma\alpha} E_d d^\dagger_{i\sigma\alpha} d_{i\sigma\alpha} + U \sum_{i\sigma\alpha} d^\dagger_{i\sigma\alpha} d_{i\sigma\alpha} d^\dagger_{i\bar\sigma\alpha} d_{i\bar\sigma\alpha} + \sum_{ij} \tilde{J}_{ij} \mathbf{S}_i \cdot \mathbf{S}_j \tag{S9}$$

where the first two terms are the interaction from single Fe sites, and the RKKY interaction between Fe sites is captured by the third term. Here, $d^\dagger_{i\sigma\alpha}$ is the creation operator of $d$-electrons from Fe atoms. $E_d$ is the energy of single Fe state with $e_g$-orbital filled of one electron, $U$ is the correlation energy of Fe $e_g$-orbital. $\mathbf{S}_i$ is the spin operator of local moment, and $\tilde{J}_{ij}$ is the exchange interaction between local moments. $\bar\sigma = -\sigma$.

The unique crystal structure of TiFe$_{0.7}$Cu$_{0.4}$Sb is that Fe atoms are randomly distributed (see fig. S11). As results from this, (a) the local moments $S_i$ are fluctuating from sites to sites, i.e., $S_i = a_i S$, where $a_i$ are random numbers that $a_i \in [0,1]$; and (b) the exchange interaction between local moments $\tilde{J}_{ij}$ are randomly distributed. For easier theoretical analysis, we may incorporate the fluctuating nature of $S_i$ (i.e., $a_i$) into the random distributed $\tilde{J}_{ij}$ by redefining that $J_{ij} \equiv a_i a_j \tilde{J}_{ij}$. Then, the local moments $S_i$ in eq. (S9) has regular behaviours as in periodic Kondo lattice, whereas the matrix elements of $J_{ij}$, despite having many zero elements, are still random as to counter not only the fluctuation of local moments but also the randomly distributed exchange interactions.

Most importantly, there is a hybridization term between conduction electrons and localized $d$-electrons, as

$$H_I = \sum_{ij\sigma\alpha} \left( V_{ij\sigma\alpha} d^\dagger_{i\sigma\alpha} c_{j\sigma\alpha} + V^*_{ij\sigma\alpha} c^\dagger_{j\sigma\alpha} d_{i\sigma\alpha} \right) \tag{S10}$$

where $V_{ij\sigma\alpha}$ is the hybridization constants.

Therefore, the total Hamiltonian of TiFe$_x$Cu$_{2x-1}$Sb is then

$$H = \sum_{ij\sigma\alpha} t_{ij} c^\dagger_{i\sigma\alpha} c_{j\sigma\alpha} + \sum_{ij} J_{ij} S_i S_j$$
$$+ \sum_{i\sigma\alpha} E_d d^\dagger_{i\sigma\alpha} d_{i\sigma\alpha} + U \sum_{i\sigma\alpha} d^\dagger_{i\sigma\alpha} d_{i\sigma\alpha} d^\dagger_{i\bar{\sigma}\alpha} d_{i\bar{\sigma}\alpha} + \sum_{ij\sigma\alpha} \left( V_{ij\sigma\alpha} d^\dagger_{i\sigma\alpha} c_{j\sigma\alpha} + V^*_{ij\sigma\alpha} c^\dagger_{j\sigma\alpha} d_{i\sigma\alpha} \right) \tag{S11}$$

where the three terms in the second line composes of an Anderson impurity model. Taking the hybridization term involving $V_{ij\sigma\alpha}$ and $V^*_{ij\sigma\alpha}$ as perturbations, the three occupied states which can span the space are $|d^1\rangle$, $|d^0\rangle$ and $|d^2\rangle$, with energies of $E_f$, $E_f + E_d$, and $E_f + 2E_d + U$.

At first, we can ignore the double occupied $|d^2\rangle$ state. Compared with $|d^1\rangle$ or $|d^0\rangle$ state, the energy level of $|d^2\rangle$ state is very large due to the presence of $U$ that typically around 1 eV (where $E_d$ is around –0.1 eV). So, the space is only spanned by state $|d^1\rangle$ and $|d^0\rangle$. Introducing the Hubbard operator $d^\dagger_\sigma = X_{\sigma 0} = |d^1, \sigma\rangle\langle d^0|$; $d_\sigma = X_{0\sigma} = |d^0\rangle\langle d^1, \sigma|$; and $d^\dagger_\sigma d_\sigma = X_{\sigma\sigma} = |d^1, \sigma\rangle\langle d^1, \sigma|$, eq. S11 is renormalized as:

$$H = \sum_{ij\sigma\alpha} t_{ij} c^\dagger_{i\sigma\alpha} c_{j\sigma\alpha} + \sum_{ij} J_{ij} S_i S_j$$
$$+ E_d \sum_{i\sigma\alpha} X^{i\alpha}_{\sigma\sigma} + \sum_{ij\sigma\alpha} \left( V_{ij\sigma\alpha} X^{i\alpha}_{\sigma 0} c_{j\sigma\alpha} + V^*_{ij\sigma\alpha} c^\dagger_{j\sigma\alpha} X^{i\alpha}_{0\sigma} \right) \tag{S12}$$

where the double occupied states (the term relating to $U$) is projected out.

In the present research, we explore the QCP physics at $T \to 0$ limit, and hence the low-energy Hilbert space is only involving the $|d^1\rangle$ state. The residual $|d^1\rangle$ state still interact with the surrounding conduction sea for virtual charge fluctuations via the process: $|d^1, \sigma\rangle + e_{\bar{\sigma}} \leftrightarrow e_{\bar{\sigma}} + e_\sigma \leftrightarrow |d^1, \bar{\sigma}\rangle + e_\sigma$ with energy $-E_d$ ($e_\sigma$, conduction electron with spin $\sigma$), which is the only process

that we take into account. From second-order perturbation theory, this virtual charge fluctuation will lower the energy by an amount of order $\Delta E = -J_K$, where $J_K = V^2/E_d$. The reduction in the energy constitutes an effective antiferromagnetic interaction between the conduction electrons and the local moments. Introducing the operator $\sigma(0) = \sum_{i,j} c_{i\alpha}^\dagger \sigma_{\alpha\beta} c_{j\beta}$, which measures the electron spin at the origin ($\sigma_{\alpha\beta}$ is the Pauli matrix), the effective interaction between the conduction electrons and $d$-electrons will have the form $H_{eff} = J_K \mathbf{S} \cdot \sigma(0)$. Thus, eq. S12 is rewritten as:

$$H = \sum_{ij\sigma\alpha} t_{ij} c_{i\sigma\alpha}^\dagger c_{j\sigma\alpha} + \sum_{i<j} J_{ij} \mathbf{S}_i \cdot \mathbf{S}_j + J_K \sum_{i\sigma\sigma'\alpha} \mathbf{S}_i \cdot \left(c_{i\sigma\alpha}^\dagger \boldsymbol{\sigma}_{\sigma\sigma'} c_{i\sigma'\alpha}\right) \quad (S13)$$

which is the so called random-exchange Heisenberg-Kondo Hamiltonian.

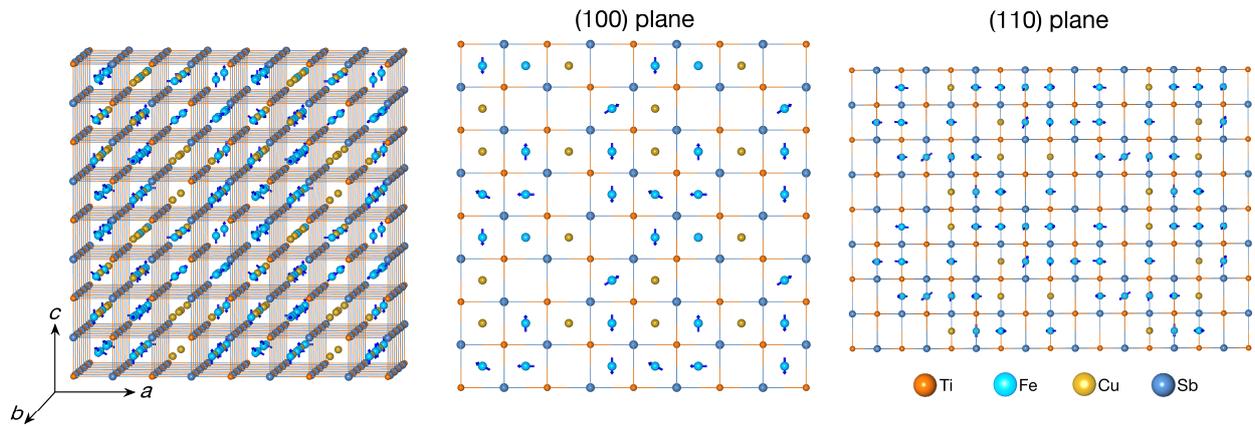

**fig. S11: Supercell crystal structure of TiFe$_{0.7}$Cu$_{0.4}$Sb.** Left, The Ti and Sb atoms form the skeleton of a face-centered cubic structure, while Fe/Cu atoms partially occupy 4$c$ and 4$d$ Wyckoff sites at random. The local magnetic moments, primarily located on the Fe atoms, are randomly distributed, with nearest neighboring antiferromagnetic interactions. The middle and right panels show views from (100) plane and (110) plane, respectively. The arrows on the Fe atoms are intended to indicate the possible presence of magnetic moments but do not represent the magnitude of these moments.

**11. Fitting of the magnetic Grüneisen parameter**

In the main text, we fit the magnetic Grüneisen parameter using $\Gamma_B(\mu_0 H) = A/(\mu_0 H - \mu_0 H_c)^{vz}$ at $\mu_0 H > 0.14$ T regime by fixing $\mu_0 H_c = 0.13$ T as obtained from the specific heat measurements. In fig. S12, we have fitted the data point to the formula in different conditions.

We also fitted the $\Gamma_B(\mu_0 H)$ by setting all parameters free, as shown in fig. S12B, but the obtain parameters, especially $\mu_0 H_c$ has a very large uncertainty, ranging from -0.108 T to 0.062 T that covers the unphysical negative regimes. For AFM or FM materials possessing Hertz-Millis type of QCPs, the critical exponent $vz$ is 1 or 3/2, respectively. In fig. S12C and D, we fitted the $\Gamma_B(\mu_0 H)$ by fixing $vz =1$ and 3/2, and the obtained critical magnetic field $\mu_0 H_c$ is also a unphysical negative value, which indicates the QCP as observed here is not conform to the Hertz-Millis type of QCPs as in AFM or FM quantum critical materials.

Because of the large noise in our measurements due to the small magnetic moments and for a fair report, we fixed $\mu_0 H_c = 0.13$ T in the fitting as shown in the main text (Fig. 2E). A complete characterization of the critical behaviours requires future experimental as well as theoretical work to refine and understand the results.

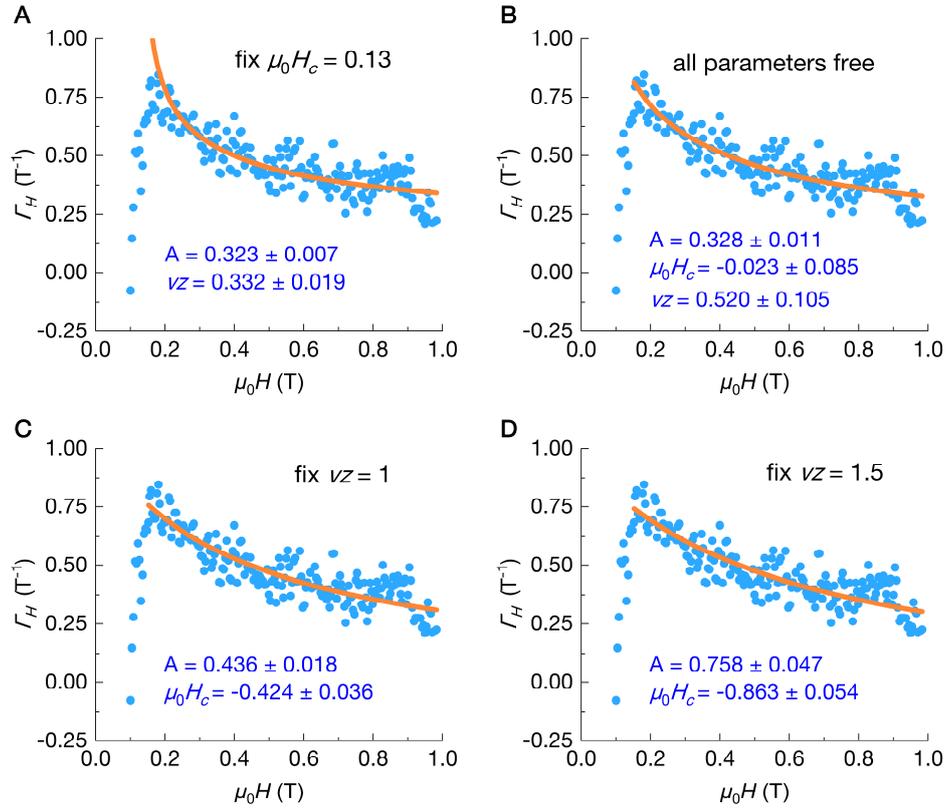

**fig. S12: Fitting of the magnetic Grüneisen parameter. A**, The magnetic Grüneisen parameter, $\Gamma_B(\mu_0H)$, is fitted by fixing $\mu_0H_c = 0.13$ T, with the fitted values indicated in blue. **B**, A direct fit without fixed parameters shows large uncertainty in $\mu_0H_c$. **C**, $\Gamma_B(\mu_0H)$ is fitted by fixing the critical exponent $vz = 1$. **D**, The same fitting process as in **C** but with $vz = 1.5$. The negative critical field values suggest the observed quantum critical point does not conform to the Hertz-Millis type in antiferromagnetic or ferromagnetic systems.

## 12. Estimation of the Kondo temperature

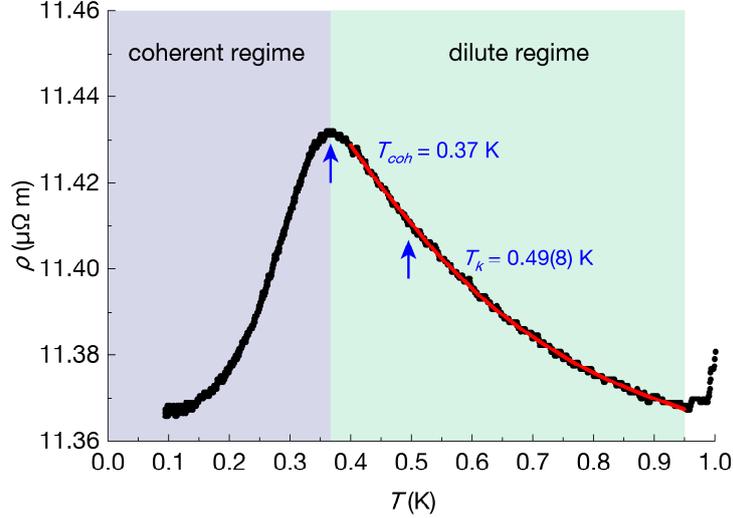

**fig. S13: Estimation of the Kondo temperature.** The resistivity $\rho(T)$ at $\mu_o H = 0$ T of TiFe$_{0.7}$Cu$_{0.4}$Sb has the similar behaviour to other dilute Kondo disordered systems. Above 0.95 K, it displays metallic behaviour (see fig. S4A as well), followed by a resistivity minimum at 0.95 K. Below this temperature, the resistivity increases with cooling (following a $-\ln T$ dependence), indicating dilute single-ion Kondo scattering (49). As the temperature is further decreased, the coherent Kondo interaction develops, featuring that $\rho(T)$ decreases with cooling, and the resistivity maximum can be intuitively defined as the coherent temperature $T_{coh}$ (see details in main text). To estimate the Kondo temperature $T_K$, we use the Hamann's expression $\rho(T) = \rho_0 + C\left\{1 - \ln\left(\frac{T}{T_K}\right)\Big/\sqrt{\ln^2\left(\frac{T}{T_K}\right) + D}\right\}$ to fit $\rho(T)$ in the dilute regime (50). The red line is the fit in the temperature region from 0.40 K to 0.95 K, and the obtained Kondo temperature $T_K$ is $0.498 \pm 0.004$ K.

## 13. Crossover between coherent and dilute regimes

As elaborated in the main text, the high-temperature electrical resistivity $\rho(T)$ is typical of uncorrelated, paramagnetic local moments in the presence of single-ion impurity Kondo hybridization with the conduction electrons, which is responsible for the negative slope. At temperatures below a crossover marked by maximal resistivity $\rho(T)$, the Kondo hybridization yields coherent electronic bands, resulting in a metallic temperature-dependence of the resistivity $\rho(T)$ due to the emerging of coherent Kondo interactions.

From the resistivity measurements, we have determined the regimes separating coherent and dilute Kondo regimes marked by the coherent temperature scale $T_{coh}$, which is determined as the maximum point at the resistivity $\rho(T)$, as shown in fig. S14A as red arrows.

Moreover, when the system transitions from the dilute Kondo impurity scattering to Kondo coherent scattering, the electronic behaviours are changed, which affects the magnetoresistivity as well. From the magnetoresistivity $\rho(\mu_0 H)$ measurmenets, we observed the $\rho(\mu_0 H)$ has a crossover from larger slope to a small slope with increasing the magnetic fields $\mu_0 H$, while the coherent magnetic field scale $\mu_0 H_{coh}$ is denoted by the blue arrows in fig. S14B.

In fig. S14C, we plotted $T_{coh}$ and $\mu_0 H_{coh}$, which coincides together separating the Kondo coherent and dilute regimes. The phase diagram reveals that increasing magnetic fields suppresses the Kondo coherent regime. This observation aligns with the Kondo breakdown scenario, where applying magnetic fields enhances Kondo screening, effectively reducing the number of active local moments as they form composite fermions with conduction electrons. Consequently, the coherent interactions between magnetic moments are gradually suppressed as more magnetic moments become screened.

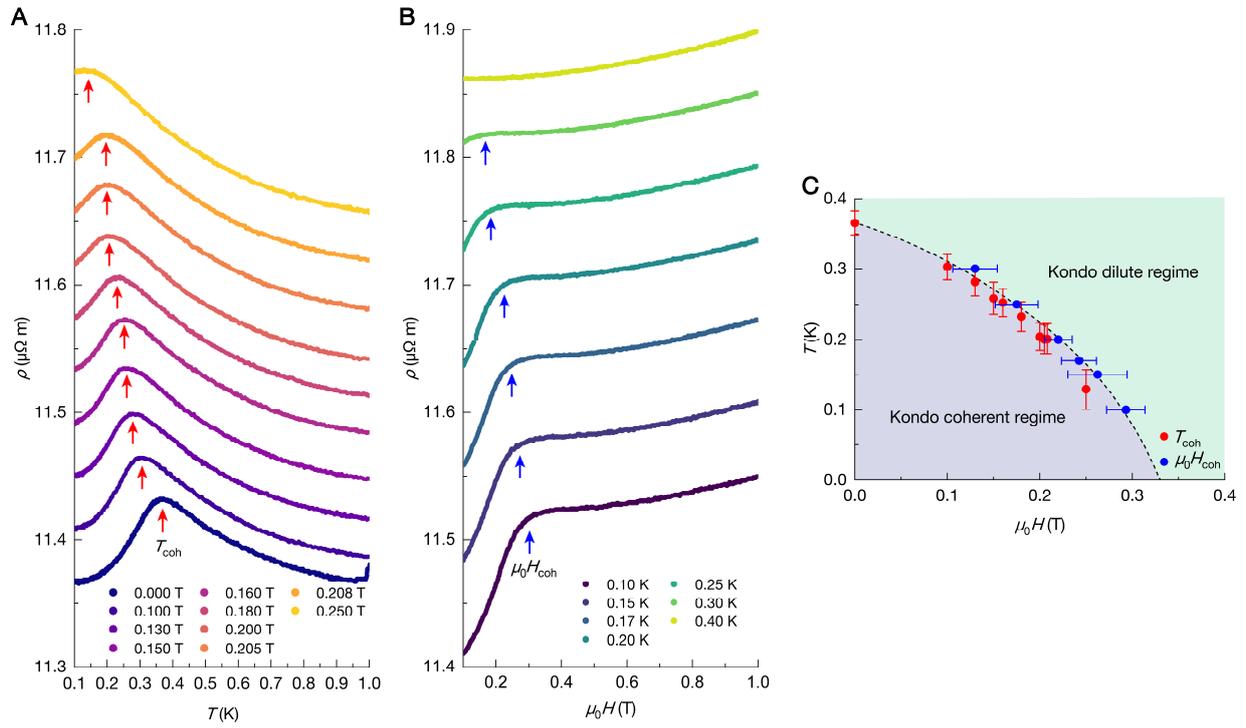

**fig. S14: Phase diagram of Kondo coherent and Kondo dilute regimes.** **A**, the resistivity $\rho(T)$ measurements, and the temperature scale $T_{coh}$ is marked by the red arrows. **B**, the magnetoresistivity measurements, and the magnetic field scale $\mu_0 H_{coh}$ is marked by blue arrows. **C**, the phase diagram which separates the Kondo coherent and dilute regimes.

## 14. ac and dc magnetic susceptibility at finite magnetic fields

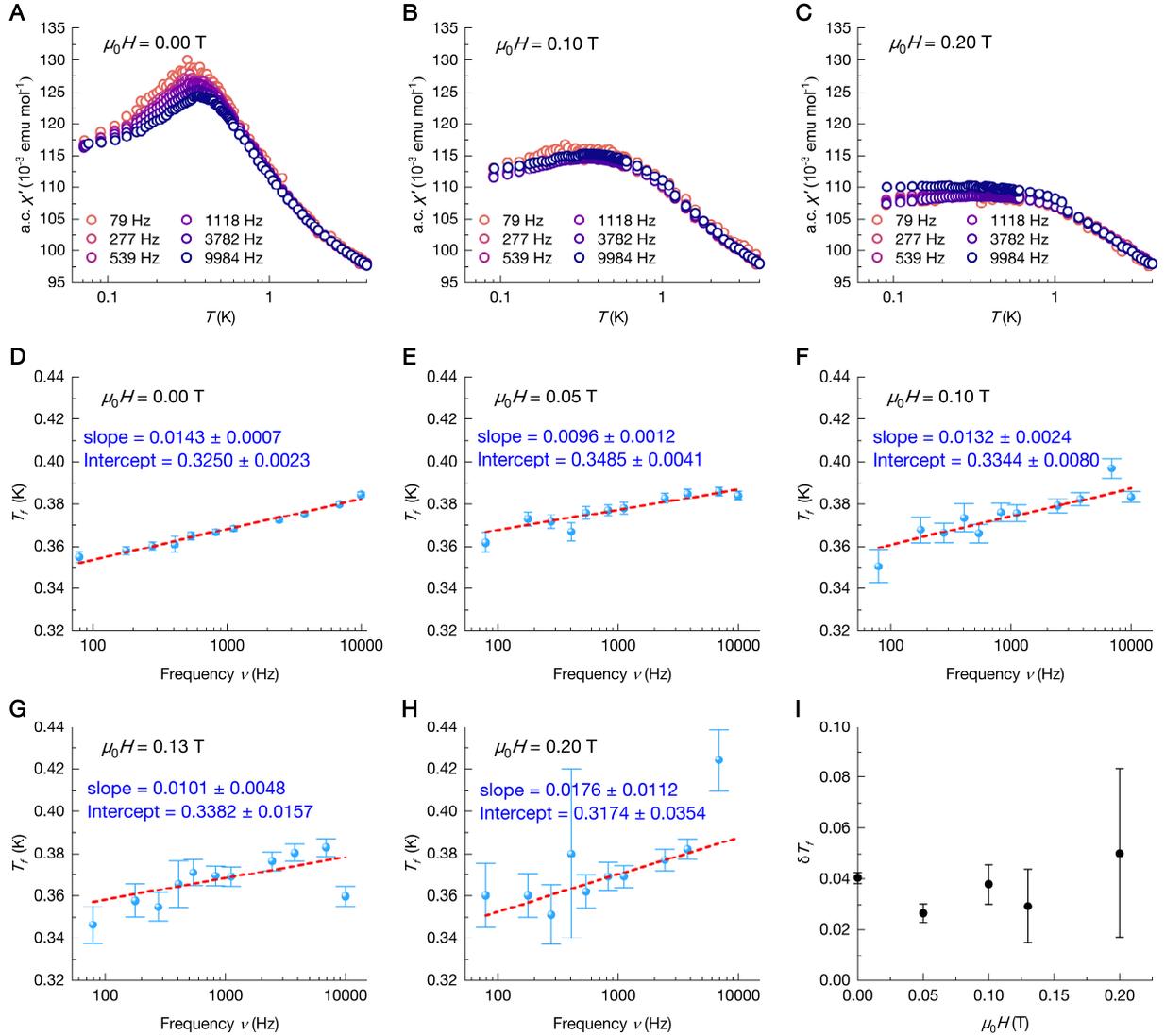

**fig. S15: ac susceptibility $\chi'(T)$ at finite magnetic fields.** With increasing the magnetic fields, the peaks in $\chi'(T)$ are gradually suppressed and the peaks height do not monotonically decrease with increasing the frequency as shown in **A**, **B** and **C**, indicating the cluster spin-glass phase is gradually diminishing and goes to the heavy-fermion liquid phase. Due to the suppression of the peaks, we cannot determine the position of $T_f$ at finite magnetic fields preciously, as reflected in the gradually increased error bars from **D** to **H**. We also tried to calculate the $\delta T_f$, as shown in **I**, which also shows significant error bars at large magnetic fields. Thus, a quantitative description of the suppression of cluster spin-glass phase cannot be given, but a qualitative description that the cluster spin-glass phase is suppressed with increasing fields can certain be concluded.

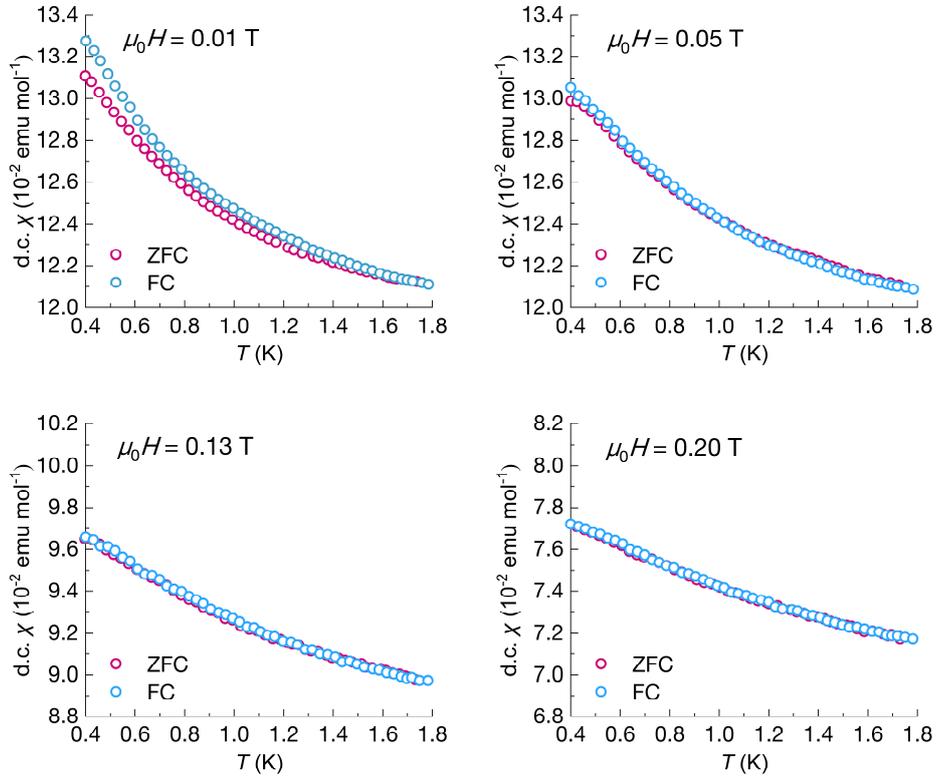

**fig. S16: dc susceptibility $\chi(T)$ at finite magnetic fields.** As the magnetic field increases, we observe that the FC and ZFC curves gradually level off, and the difference between them diminishes at 0.13 T. These features indicate that the cluster spin-glass phase at $T < 0.4$ K is progressively suppressed as the magnetic field strengthens.

## 15. Quantitative analysis of the crossover functions in Hall measurements

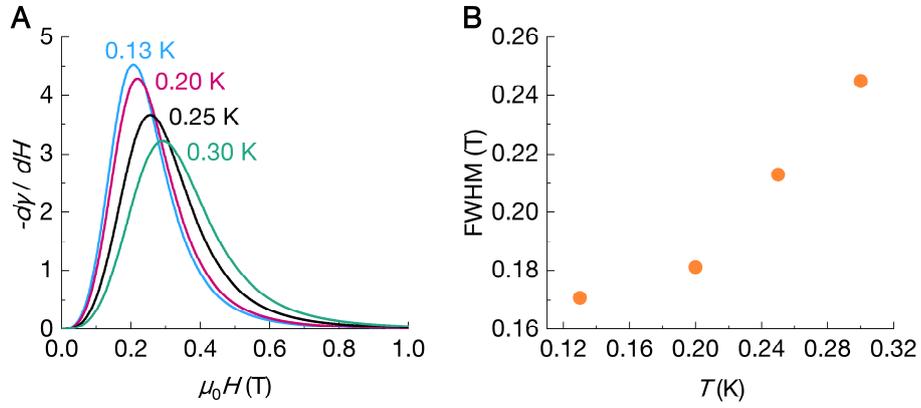

**fig. S17: Quantitative analysis of the crossover function $\gamma(\mu_0 H)$.** As described in the main text, we used a crossover function $\gamma(\mu_0 H)$ to fit the Hall resistivity. In panel **A**, we plot the field derivative of the crossover function, which can be interpreted as the rate of change of the effective carrier concentration. As the temperature approaches zero, the crossover function is expected to diverge at the QCP. Panel **B** shows the full width at half maximum (FWHM), indicating the width of the crossover. With decreasing temperature, the FWHM decreases, consistent with the sharpening expected for quantum criticality, as observed in other materials exhibiting a QCP.

## 16. Kadowaki-Woods relation

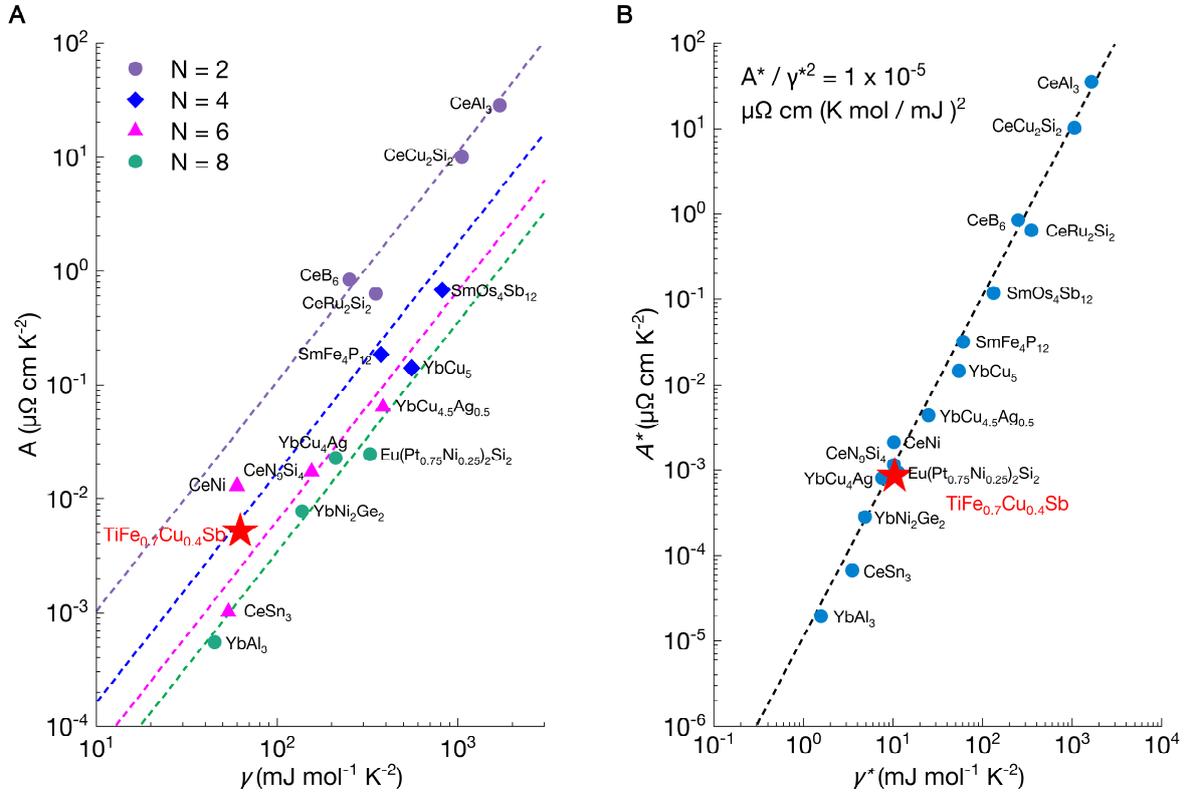

**fig. S18. Kadowaki-Woods relation in TiFe$_{0.7}$Cu$_{0.4}$Sb.** The Sommerfeld coefficient $\gamma$ obtained at $\mu_0 H = 0$ T is 62.55 mJ mole-Fe$^{-1}$ K$^{-2}$. The coefficient $A$ is obtained by fitting the resistivity at $\mu_0 H = 0$ T to $\rho(T) = \rho_0 + AT^2$ below $T = 0.30$ K, yielding $A = 0.51489 \pm 0.00211 \times 10^{-2}$ Ω cm K$^{-2}$. The position of TiFe$_{0.7}$Cu$_{0.4}$Sb in the $A$-$\gamma$ diagram is plotted in **A**, along with the data of other heavy-fermion materials as classified by their spin degeneracy $N$. The dashed lines are the prediction from the orbitally degenerate periodic-Anderson model (*51, 52*). In TiFe$_{0.7}$Cu$_{0.4}$Sb, $N = 4$ due to the two-fold degeneracy of $e_g$ orbital. When the spin degeneracy is taken into consideration, $A^* = A/\frac{1}{2}N(N-1)$, and $\gamma^* = \gamma/\frac{1}{2}N(N-1)$ could be defined. The Kadowaki-Woods quantity, as defined by $A^*/\gamma^{*2}$ has been theoretically derived to be a constant value of $1 \times 10^{-5}$ μΩ cm(K mol / mJ)$^2$ (black dashed line in **B**) (*53*), which is the found hold in many heavy-fermion materials. We found TiFe$_{0.7}$Cu$_{0.4}$Sb also following this relation. The experimental data are collected from Ref. (*52, 54-57*).

## 17. Schematic of global phase diagram

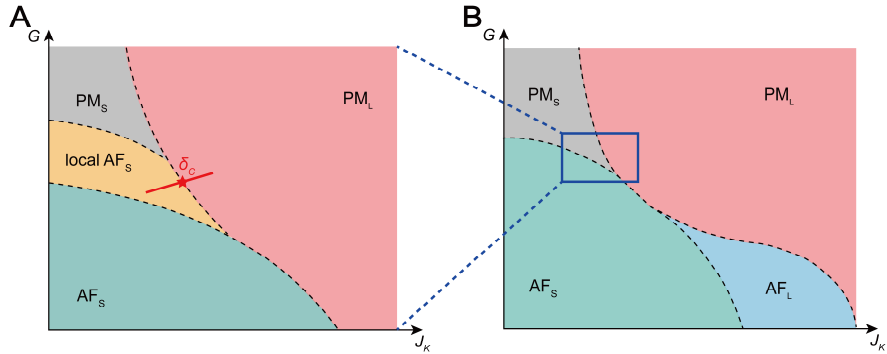

**fig. S19: Schematic of global phase diagram. A**, The modified phase diagram at the area denoted by blue rectangle in **B**, which is suitable for TiFe$_{0.7}$Cu$_{0.4}$Sb. $J_K$ tunes the Kondo coupling and $G$ varies the degree of magnetic frustration. PM and AF stand for paramagnetic and antiferromagnetic phases, and the subscripts L and S denote the large and small Fermi surface. Since the cluster spin glass contains local antiferromagnetic pairs without long-range magnetic order, generally speaking, it is not an AF or PM states, but a state in between denoted by local AF$_s$. The red line denotes the path for phase transition, and $\delta_c$ is the point of Kondo breakdown QCP, where $\delta$ is the turning parameter such as magnetic field here. **B**, Normal global phase diagram for heavy-fermion metals *(58, 59)*.

**Table S1: Parameters for substruction of the phonon $C_{ph}$, nuclear $C_{nuc}$ and magnetic Schottky contribution $C_{sch}$ to the total specific heat.** For phonon contributions ($C_{ph} = \beta T^3$), we only fitted the data at zero field, and the same $\beta$ was employed for finite fields. For nuclear contributions ($C_{nuc} = A/T^2$), we fitted the data at all magnetic fields at $T < 0.1$ K region. For magnetic Schottky contributions, $n$ and $\Delta$ are determined by the peak height and position of $C_m$, respectively. Note that at these magnetic fields of 0.025, 0.075, 0.140 and 0.250 T, specific heat is only measured below 0.15 K. N/A: not applicable.

| $\mu_0 H$ (T) | $\beta$ (J mol$^{-1}$ K$^2$) | $A$ (10$^{-5}$ J mol$^{-1}$ K) | $n$ (%) | $\Delta$ (K) |
| --- | --- | --- | --- | --- |
| 0 | 0.00105±0.00006 | 1.93488±0.02384 | 0.28 | 1.16 |
| 0.025 | / | 1.90843±0.02659 | N/A | N/A |
| 0.050 | / | 1.91530±0.03160 | 0.27 | 1.20 |
| 0.075 | / | 1.84519±0.03298 | N/A | N/A |
| 0.100 | / | 1.86741±0.05257 | 0.26 | 1.24 |
| 0.125 | / | 1.86359±0.03050 | 0.26 | 1.26 |
| 0.130 | / | 1.83464±0.02310 | 0.27 | 1.27 |
| 0.140 | / | 1.87062±0.02728 | N/A | N/A |
| 0.150 | / | 1.89023±0.03213 | 0.26 | 1.31 |
| 0.175 | / | 1.77218±0.02908 | 0.26 | 1.36 |
| 0.200 | / | 1.81643±0.02648 | 0.26 | 1.42 |
| 0.250 | / | 1.79424±0.02653 | N/A | N/A |
| 0.300 | / | 1.83215±0.03908 | 0.32 | 1.58 |
| 0.400 | / | 1.92868±0.04562 | 0.37 | 1.75 |
| 0.500 | / | 1.95660±0.04644 | 0.41 | 1.93 |
| 0.600 | / | 2.00549±0.08831 | 0.45 | 2.11 |